\documentclass[12pt,a4paper]{article}
\usepackage{graphicx}
\usepackage{amssymb}
\usepackage{amsmath}
\usepackage{bm}
\usepackage{color}
\usepackage{theorem}
\usepackage{subfigure}
\usepackage{caption}
\usepackage{todonotes}
\usepackage{physics}

\usepackage[sort&compress,numbers, merge]{natbib}

\definecolor{Orange}{cmyk}{0,0.61,0.87,0}
\definecolor{JungleGreen}{cmyk}{0.99,0,0.52,0}
\definecolor{OliveGreen}{cmyk}{0.64,0,0.95,0.40}
\definecolor{Brown}{cmyk}{0,0.81,1,0.60}
\definecolor{RoyalBlue}{cmyk}{0.71,0.53,0,0.12}

\allowdisplaybreaks[1]

\newcommand{\Slash}[1]{{\ooalign{\hfil/\hfil\crcr$#1$}}}

\usepackage[colorlinks=true, linkcolor=OliveGreen, citecolor=RoyalBlue,
urlcolor=RoyalBlue]{hyperref}

\begin{document}

\begin{titlepage}
\pagestyle{empty}

{\tt \rightline{IPMU20-0102}}
\vspace{1cm}
\begin{center}
{\bf {\LARGE
Proton Decay in Product Group Unification} }
\end{center}

\vspace{0.05cm}
\begin{center}
{\bf Jason~L.~Evans}~$^1$,
 {\bf Masahiro~Ibe}~$^{2,3}$,
{\bf Tsutomu~T.~Yanagida}~$^{1,3}$
\vskip 0.2in
{\small {\it
$^1${T. D. Lee Institute, Shanghai 200240, China}\\
\vspace{0.2cm}
$^2${Institute for Cosmic Ray Research (ICRR), The University of Tokyo, Chiba 277-8583, Japan}\\
\vspace{0.2cm}
$^3$ {Kavli Institute for the Physics and Mathematics of the Universe
 (WPI), \\The University of Tokyo Institutes for Advanced Study, \\ The
 University of Tokyo, Kashiwa 277-8583, Japan}
}

}

\vspace{0.5cm}

{\bf Abstract}\\
\end{center}
Product group unification is an
attractive alternative to simple grand unification.
It solves the infamous doublet-triplet splitting problem and the dimension-5 proton decay problems without introducing any fine-tuning. Furthermore, the matter multiplets are still embedded into unified SU(5) representations. In this paper, we discuss proton decay of the simplest product group unification model based on SU(5)$\times$U(2)$_\mathrm{H}$.
We find that the minimal setup of the model has already been excluded by dimension-6 proton decay.
We also show that a simple extension of the model, with naturally generated SU(5) incomplete multiplets, can rectify this problem.
We find that the proton lifetime will be in reach of coming experiments like DUNE and Hyper-K, when the mass of the incomplete multiplet is associated with the Peccei-Quinn symmetry breaking. In this case, the dark matter may be an admixture of the Wino LSP and the axion.

{}


\vfill
\end{titlepage}

\section{Introduction}
One of the challenges of supersymmetric (SUSY) grand unified theories (GUT), is the doublet-triplet splitting.
In the minimal model of SU(5), for example, the Higgs bosons must be embedded in a $\mathbf{5},\mathbf{\bar 5}$.
This means that the doublet Higgs bosons are accompanied by SU(3) triplets.
The presence of a triplet Higgs boson,
at the energy scale of the
minimal supersymmetric standard model (MSSM), ruins the precise coupling unification at the GUT scale.
In addition to this complication, the predicted lifetime for the proton through the exchange of this very light triplet Higgs boson would be in conflict with experimental constraints.  Thus, the doublet-triplet splitting is required. In minimal SU(5), this splitting is accomplished by a severe fine tuning.

The minimal SU(5) model is further complicated by the fact that the proton, lifetime in the channel $p\to K^+ \bar\nu$, tends to be too short unless the soft masses are quite large and the phases in the Yukawa couplings are chosen appropriately\footnote{Minimal SU(5) models are further complicated by the fact that the operator ${\bf \bar 510 10 10}/M_P$ is allowed by all the symmetries. Unless the coefficient of this operator is quite small, the short proton lifetime rules out all low-scale SUSY models.} ~\cite{Sakai:1981pk,Weinberg:1981wj} (for recent work, see e.g.~\cite{Ellis:2016tjc,Ellis:2017djk,Ellis:2019fwf}). This has lead to the study of more sophisticated model.

Several solutions to the doublet-triplet splitting have been proposed.
One candidate, Missing partner models~\cite{Masiero:1982fe,Grinstein:1982um}, use a $\mathbf{75}$ to break SU(5) down to the standard model (SM) gauge symmetries. In this case, the Higgs bosons, which reside in a $\mathbf{5},\mathbf{\bar 5}$, are coupled a $\mathbf{50},\mathbf{\overline{50}}$ through the $\mathbf{75}$. Since the $\mathbf{50},\mathbf{\overline{50}}$ do not contain any doublets, the Higgs bosons bilinear mass is not generated.
Although this model solves the doublet-triplet splitting problem, a more complicated structure is needed to forbid the $\mathbf{5}\mathbf{\bar 5}$ Higgs bilinear term. This structure tends to be plagued by other problems. Other types of unification models, like flipped SU(5), also rely on a missing partner type mechanism to suppress the Higgs doublet mass. However, the $\mathbf{5\bar 5}$ Higgs bilinear mass term is set to zero by hand\footnote{It is possible using R-symmetries to forbid the Higgs bilinear mass term in flipped SU(5), see~\cite{Hamaguchi:2020tet}}.  Furthermore, models like flipped SU(5) completely lose the explanation of charge quantization.

In this paper, we will examine product group unification with the gauge symmetries SU(5)$\times$ U(2)$_{\mathrm{H}}$~\cite{Ibe:2003ys,Izawa:1997he} (see Refs.~\cite{Yanagida:1994vq,Hisano:1995hc,Hotta:1996qb,Hotta:1996pn,Izawa:1997he} for the earlier works on this type of the product group unification models.).
This product group unification model is characterized by having an R-symmetry
which forbids
the Higgs bilinear term $\mathbf{5\bar{5}}$ and the dimension-five proton decay operators simultaneously\footnote{This also forbids the operator ${\bf 5 10 10 10}/M_P$.}.
For this model, the doublet-triplet splitting is accomplished without any unnaturally small couplings.

Furthermore, this model maintains the same matter field embeddings as minimal SU(5), i.e., the standard model fields are contained in the $\mathbf{10}$ and $\mathbf{\bar 5}$ just as in minimal SU(5).
Although the SM gauge symmetries SU(2)$\times$U(1) are the diagonal subgroup of SU(5)$\times$U(2)$_\mathrm{H}$, this embedding leads to a perceived charge quantization among the MSSM fields due to the embedding of SM fields in the $\mathbf{10}$ and $\mathbf{\bar 5}$\footnote{However, any charge is possible if the $\mathbf{10}$'s and $\mathbf{\bar 5}$'s are initially charged under the U(1)$_\mathbf{H}$ or there are other fields charged only under the U(2)$_\mathrm{H}$ gauge symmetries.}.

As we will see, the minimal SU(5)$\times$ U(2)$_{\mathrm{H}}$ unification model has already been excluded by dimension-6 proton decay experiments\footnote{This is due the gauge coupling matching conditions requiring the SU(5) guage bosons to be light.}~\cite{Miura:2016krn}.  We also show that a simple extension of the model including new pairs of $\mathbf{5}$, $\mathbf{\bar 5}$ and $\mathbf{2}$, $\mathbf{\bar 2}$ can rectify this problem. In this extension, the mass of the doublets and the triplets embedded in the new $\mathbf{5},\mathbf{\bar 5}$ are split with the triplets being much lighter. These light triplets deflect the running of the gauge couplings and alters the gauge matching conditions. This leads to a larger mass for the heavy gauge bosons of SU(5) and a longer proton lifetime. As we will show, these light triplets can also be the heavy quarks which couple to the Peccei-Quinn breaking field of the KSVZ axion scenario~\cite{Kim:1979if,Shifman:1979if}.
To push the proton lifetime beyond the current experimental limit, the triplet masses need to be smaller than about $10^{12}$ GeV.
Suggestively, this scale implies that the axion makes up some portion of the dark matter. If the axion does indeed make of some non-trivial portion of the dark matter and we make some rather mild assumptions about order one couplings, the proton lifetime of this model will be in reach of coming experiments like DUNE~\cite{Abi:2018dnh} and Hyper-K~\cite{Abe:2018uyc} no matter the MSSM soft mass spectrum. Since dimension-6 proton decay dominates, this makes for a unique proton decay signature for these experiments to search for.

The paper is organized as follow. Section 2 gives a brief review of SU(5)$\times$ U(2)$_{\mathrm{H}}$ production group unification. Next, in section 3, we discuss our calculation of the proton lifetime including discussion of the SUSY breaking scenario we use. In section 4, we presents our product group unification model, including some discussion on how the axion ties in to this scenario. Then, section 5 presents the result of our proton lifetime calculation for the particular product group unification model we consider.

\section{The Model}
The model we consider is based on that found in \cite{Izawa:1997he,Ibe:2003ys} and is a unification model with the gauge symmetries SU(5)$\times$ U(2)$_{\mathrm{H}}$. These symmetries are broken down to the SM gauge symmetries SU(3)$\times$SU(2)$\times$U(1) by the following superpotential
\begin{eqnarray}
&W=\sqrt{2}\lambda_{2H}\bar Q X^a\tau^a Q +\sqrt{2}\lambda_{1H}\bar Q X_0 Q
-\sqrt{2}\lambda_{1H}v^2 X_0\ ,
\end{eqnarray}
where we have suppressed gauge indices and the charge assignments are in Table~\ref{table1}.
The $\tau^a$ $(a=1,2,3)$ denote half of the Pauli matrices, $v$ is the
mass parameter of the GUT scale, while $\lambda$'s are coupling constants.
We follow the normalization of the coupling constants in \cite{Ibe:2003ys}.
\begin{table*}[t]
\caption{Charge assignments for all fields.
We normalize the U(1)$_\mathrm{H}$ charge so that the charge matrix on the SU(2)$_\mathrm{H}$ doublet is $(\tau_0)_{\alpha\beta} = \delta_{\alpha\beta}/2$.
}
\begin{center}
\resizebox{\textwidth}{!}{
\begin{tabular}{|c|c|c|c|c|c|c|c|c|c|c|c|c|c|c|c|c|}
\hline  Fields &  ~~$\bar\Phi_i$~~ & ~~$\Psi_i$~~&~~$X$~~&~~ $X_0$~~ & ~~$Q_6$~~ &~~ $ \bar Q_6$~~&~~$Q$~~&~~$\bar Q$~~&~~$\Phi'$~~& ~~ $\bar \Phi'$ ~~&~~ $\Theta$~~ & ~~ $\bar \Theta$~~ &~~$P$~~\\
\hline SU(5) & $\mathbf{\bar 5}$ & $\mathbf{10}$ & $\mathbf{1}$ &$\mathbf{1}$& $\mathbf{1}$ & $\mathbf{1}$ & $\mathbf{\bar 5}$ & $\mathbf{5}$ & $\mathbf{5}$ & $\mathbf{\bar 5}$ &$\mathbf{1}$&$\mathbf{1}$&$\mathbf{1}$  \\
\hline SU(2)$_\mathrm{H}$ &    $\mathbf{1}$    & $\mathbf{1}$  &$\mathbf{3}$ &$\mathbf{1}$& $\mathbf{2}$ & $ \mathbf{2}$  & $ \mathbf{2}$ & $\mathbf{2}$& $\mathbf{1}$&$\mathbf{1}$&$\mathbf{2}$&$\mathbf{2}$ &$\mathbf{1}$\\
\hline U(1)$_\mathrm{H}$ &   0 &  0  & 0&0  & -1/2 & 1/2 &  -1/2  & 1/2 &0&0&-1/2&1/2&0 \\
\hline
R-charge &   1 &  1  & 2&2  & 0  & 0 &  0  & 0&1&1&1&1&0  \\
\hline
PQ-charge & 0 &0 &0 &0 &0 &0&0&0& 1&0&0&-1&-1\\
\hline
\end{tabular}}
\end{center}
\label{table1}
\end{table*}%
The theory also has a well defined $R$ symmetry as seen in Table \ref{table1}.

The theory is broken to the SM gauge symmetries by the vacuum expectation value (VEV)
\begin{eqnarray}
Q_\alpha^A=v\delta^A_\alpha \quad \quad Q_A^\alpha= v\delta_A^\alpha
\end{eqnarray}
where $\alpha,\beta...$ are for the SU(2) indices and $A,B..$ refer to the SU(5) indices.  After the gauge symmetry is broken, the masses of the particles are
\begin{eqnarray}
M_{X'}=\sqrt{2}\lambda_{2H}v \quad \quad M_{Q^\alpha_\beta+\bar Q^\alpha_\beta}=\sqrt{2}\lambda_{2H}v \quad \quad M_{X_0}=\sqrt{2}\lambda_{1H}v
\end{eqnarray}
where $M_{Q^\alpha_\beta+\bar Q^\alpha_\beta}$ is the mass of the linear combination $Q^\alpha_\beta+\bar Q^\alpha_\beta$ and the others we hope are self explanatory. The orthogonal component $\bar Q^\alpha_\beta-Q^\alpha_\beta$ is one of the Goldstone boson fields associated with the breaking of the gauge symmetries. The other component of $Q,\bar Q$, involving SU(3) portion of the SU(5) indices, are also would-be Goldstone bosons.
The gauge boson masses corresponding to the broken generators are as follows
\begin{eqnarray}
M_X=g_5v\ ,\quad \quad M_{V_{\mathrm{U(1)}}}=\sqrt{2}\sqrt{g_{1H}^2+\frac{3}{5}g_5^2} v \ , \quad \quad M_{V_{\mathrm{SU(2)}}}=\sqrt{2}\sqrt{g_{2H}^2+g_5^2}v~.
\end{eqnarray}
It should be noted that the SU(5)$\times$ U(2)$_{\mathrm{H}}$ breaking sector leaves no massless particles.

In this theory, we break the SU(2)$\times$ U(1) subgroup of SU(5) diagonally with the U(2)$_\mathrm{H}$.
This means that after the breaking, the U(2)$_\mathrm{H}$ fields now have SU(2)$_\mathrm{W}\times$U(1)$_\mathrm{Y}$ charges. The charges of the massive guage bosons are, in the notation $(\mathrm{SU(3)},\mathrm{SU(2)})_{\mathrm{U(1)}_\mathrm{Y}}$, $X(\mathbf{3},\mathbf{2})_{5/3}$, $\bar X(\mathbf{\bar 3},\mathbf{2})_{-5/3}$, $V_{\mathrm{SU(2)}}(\mathbf{1},\mathbf{3})_0$, and $V_{\mathrm{U(1)}}(\mathbf{1},\mathbf{1})_0$. Using these charge assignments, we get the following matching conditions for the gauge couplings
\begin{eqnarray}
&&\frac{1}{g_3^2(M_G)}=\frac{1}{g_5^2(M_G)}+\frac{1}{2\pi^2}\ln\left(\frac{M_X}{M_G}\right)\label{eq:g3mat}\ , \\
&&\frac{1}{g_2^2(M_G)}=\frac{1}{g_5^2(M_G)}+\frac{1}{g_{2H}^2}+\frac{3}{4\pi^2}\ln\left(\frac{M_X}{M_G}\right)+\frac{1}{2\pi^2}\ln\left(\frac{M_{V_\mathrm{SU(2)}}}{M_{Q^\alpha_\beta+\bar Q^\alpha_\beta}}\right)\ ,\\
&&\frac{1}{g_{1}^2(M_G)}=\frac{1}{g_5^2(M_G)}+\frac{3}{5}\frac{1}{g_{1H}^2}+\frac{5}{4\pi^2}\ln\left(\frac{M_X}{M_G}\right)\ ,
\label{eq:g1mat}
\end{eqnarray}
where we use $M_G$ to indicate the matching scale. When we implement these matching conditions, we will use the scale at which $g_1=g_2$. From Eq.\,\eqref{eq:g3mat} and \eqref{eq:g1mat}, we can find $M_X$ in terms of $g_{1H}^2$,
\begin{eqnarray}
M_X=M_G\exp\left(\frac{4\pi^2}{3}\left[\frac{1}{g_1^2(M_G)}-\frac{1}{g_3^2(M_G)}-\frac{3}{5}\frac{1}{g_{1H}^2(M_G)}\right]\right)\label{eq:MXg1h}\ .
\end{eqnarray}
Since the couplings unify quite well in supersymmetry, the differences of the MSSM gauge couplings is quite small.  If $g_{1H}^2$ is of order $4\pi$, then $M_X$ is quite close to the unification scale.  However, if $g_{1H}^2$ is of order one, then $M_X$ is much lower than the unification scale. As we will see, this leads to a proton lifetime which is in conflict with experimental constraints.

The gauge coupling $g_5$, which is important for proton decay, can be found from Eq.\,\eqref{eq:g3mat},
\begin{eqnarray}
g_5(M_G)= \left[\frac{1}{g_3^2(M_G)}-\frac{1}{2\pi}\ln\left(\frac{M_X}{M_G}\right)\right]^{-\frac{1}{2}}~.
\end{eqnarray}
Since $M_X$ is known in terms of $g_{1H}$, both $g_5$ and $M_X$ are determined by choosing $g_{1H}$.

\subsection{MSSM Yukawa Couplings}
As a notable feature of the
SU(5)$\times$ U(2)$_{\mathrm{H}}$ unification model, there are no $\mathbf{5}$ and $\mathbf{\bar 5}$ Higgs bosons.  Instead, the Higgs bosons arise from additional massless fields charged under only U(2)$_\mathrm{H}$,
$Q_6$ and $\bar Q_6$ in Table.\,\ref{table1}\footnote{The $Q_6$ and $\bar Q_6$ are nothing but the pseudo-Nambu-Goldstone chiral multiplets in the limit of $\lambda_{2H}=\lambda_{1H}$~\cite{Hotta:1996pn}.}.
Since these fields will be charged under the SM SU(2)$\times$U(1)$_\mathrm{Y}$ once the SU(5)$\times$U(2) breaks to the SM gauge symmetries, these field can play the role of Higgs boson. At the tree-level, the Higgs bosons cannot interact with the MSSM matter content. The only allowed tree-level interactions of the Higgs bosons are
\begin{eqnarray}
W_H=\sqrt{2}\lambda^\prime_{2H}\bar Q_6X Q_6+\sqrt{2}\lambda^\prime_{1H} \bar Q_6 X_0 Q_6\label{eq:Yuk}
\end{eqnarray}
In the above expression, there is no supersymmetric mass for the $Q_6,\bar Q_6$ since their R-charge is zero.
As there is no triplet Higgs, the model is free from the doublet-triplet splitting problem by construction.
We will return to the generation of the Higgs supersymmetric bilinear mass later.

The MSSM Yukawa couplings are generated from higher dimensional operators,
\begin{eqnarray}
W_Y=\frac{c_{5_{ij}}Q\bar Q_6}{\Lambda}\Psi_i  \bar\Phi_j+\frac{c_{10_{ij}}\bar Q Q_6}{\Lambda} \Psi_i \Psi_j\ ,\label{eq:MSSMYuk}
\end{eqnarray}
where $\Lambda$ is the cut off of the theory.
The $\bar\Phi$ and $\Psi$ above contain all the MSSM fields with $i,j$ being the flavor indices (see Table\,\ref{table1}). They both also have an R-charge of 1.
To reproduce the top Yukawa coupling in the MSSM, we require that the cutoff scale $\Lambda$ is not far from $\order{\langle Q\rangle}$ but larger\footnote{The top Yukawa couplings for the models we consider below are of order $0.45-0.6$ depending on the value of $\tan\beta$ and the Higgs soft masses, which require a rather large coupling $c_{10}$.}.

Now the expressions for the Yukawa couplings require a little more careful treatment, since the SU(2)$_\mathrm{H}$ and U(1)$_\mathrm{H}$ are not asymptotically free. In this case, we identify $\Lambda$ with the confinement scale of some strong interacting ultra-violet (UV) theory. Unless the Landau-pole scale is separated form the GUT scale, $\order{v}$, by at least an order of magnitude, the expressions for the Yukawa couplings above are not well defined.

This needed separation of scales has implications for the gauge couplings. If
we enforce $\Lambda\gtrsim 4\pi\langle Q\rangle$, the gauge couplings for SU(2)$_\mathrm{H}$ and U(1)$_\mathrm{H}$ will be suppressed
at the GUT scale due to the renormalization group (RG) running. That is, even if we set them equal to $4\pi$ at the cutoff scale $\Lambda$, they will no longer be of order $4\pi$ at the GUT scale. This RG running will place an upper limit on the size of the $g_{1H}$ and $g_{2H}$ at the GUT scale which is less than $4\pi$. This will in turn affects the upper bound on the the mass of $M_X$, as seen in Eq.\,\eqref{eq:MXg1h}.
Since $M_X$ only depends on $g_{1H}$, we will focus on the RG effects on this coupling. We will only consider the one-loop RGE's, using them as a guide. The one-loop RGE for $g_{1H}$ is
\begin{eqnarray}
\frac{d g_{1H}^2}{d\ln\mu}= 6\frac{g_{1H}^4}{8\pi^2} \ .
\end{eqnarray}
The solution to this one-loop equations is
\begin{align}
g_{1H}^2(\mu)=\frac{g_{1H}^2(\Lambda)}{1-6\frac{g_{1H}^2(\Lambda)}{8\pi^2}\ln\left(\frac{\mu}{\Lambda}\right)} \ .
\end{align}
Using this equation, we can determine the maximum size of the coupling that allows a $4\pi$ separation between the Landau pole, $\Lambda$ and GUT scale, which we take to be $M_G$ throughout the rest of this work. This is roughly estimated by taking $g_{1H}^2(\Lambda)=\Lambda/\mu=4\pi$ in the above equation, which gives
\begin{eqnarray}
g_{1H}^2(\langle{Q}\rangle)=g_{1H}^2\left({\Lambda}/{4\pi}\right)=3.68\ .\label{eq:g1hlandau}
\end{eqnarray}
As we will see below, a value this small leads to a proton lifetime which is much too short in the minimal
SU(5)$\times$ U(2)$_{\mathrm{H}}$ model we have dicussed above.

\section{Proton Decay}
In product group unification, there is no dimension 5 proton decay, since the operator $\mathbf{\bar 5 10 10 10}$ is forbidden by the R-symmetry. This is already a significant deviation from minimal SU(5) where this is the dominant decay mode.

Dimension-6 proton decay, on the other hand, proceeds as usual. Here we will give some details of the dimension-6 proton decay calculation. The important interactions for dimension-6 proton decay are
\begin{align}
 {\cal L}_{\rm int} = \frac{g_5}{\sqrt{2}}
\left[
- \overline{d^c_{Ri}} \Slash{X} L_i
+ e^{-i\varphi_i}\overline{Q}_i \Slash{X} u_{Ri}^c
+ \overline{e_{Ri}^c} \Slash{X} (V^\dagger)_{ij}Q_j
+ {\rm h.c.}
\right]~,\label{eq:Xint}
\end{align}
where $V_{ij}$ are the CKM matrix elements.
To calculate the proton lifetime induced by these operators, we first integrate out the $X$ boson. We then evolve these operators Wilson's coefficients down to the hadronic scale using renormalization group equations, making the necessary adjustments to the equations at the SUSY and weak scale.  The decay width is then calculated at the hadronic scale\footnote{We take the bottom quark mass as the hadronic scale.} for different leptons flavors $\ell_i$,
\begin{eqnarray}
 \Gamma (p\to  \pi^0 \ell_i^+)=
\frac{m_p}{32\pi}\biggl(1-\frac{m_\pi^2}{m_p^2}\biggr)^2
\bigl[
\vert {\cal A}_L(p\to \pi^0 \ell_i^+) \vert^2+
\vert {\cal A}_R(p\to \pi^0 \ell_i^+) \vert^2
\bigr]~,
\end{eqnarray}
where $m_p$ and $m_\pi$ are the proton and pion masses respectively.  The amplitudes are given by
\begin{align}
 {\cal A}_L(p\to \pi^0 \ell_i^+)&=
- \frac{g_5^2}{M_X^2}\delta_{i1}\cdot
A_1 \cdot \langle \pi^0\vert (ud)_Ru_L\vert p\rangle_i
~,\nonumber \\
 {\cal A}_R(p\to \pi^0 \ell_i^+)&=
- \frac{g_5^2}{M_X^2} (\delta_{i1}+V_{ud}V_{ui}^*) \cdot
A_2 \cdot \langle \pi^0\vert (ud)_Lu_R\vert p\rangle_i
~,
\end{align}
where $A_{1,2}$ takes care of the renormalization group (RG) running, and $\langle \pi^0\vert (ud)_{(R,L)}u_{(L,R)}\vert p\rangle_i$ are the hadron matrix element for decays to $\ell_i$.  The (RG) coefficients $A_{1,2}$ are given by
\begin{align}
 A_1 =&
A_L \cdot \biggl[
\frac{\alpha_3(M_{\text{SUSY}})}{\alpha_3(M_{\rm GUT})}
\biggr]^{\frac{4}{9}}
\biggl[
\frac{\alpha_2(M_{\text{SUSY}})}{\alpha_2(M_{\rm GUT})}
\biggr]^{-\frac{3}{2}}
\biggl[
\frac{\alpha_1(M_{\text{SUSY}})}{\alpha_1(M_{\rm GUT})}
\biggr]^{-\frac{1}{18}}
\nonumber \\[2pt]
&\times
\biggl[
\frac{\alpha_3(m_Z)}{\alpha_3(M_{\rm SUSY})}
\biggr]^{\frac{2}{7}}
\biggl[
\frac{\alpha_2(m_Z)}{\alpha_2(M_{\rm SUSY})}
\biggr]^{\frac{27}{38}}
\biggl[
\frac{\alpha_1(m_Z)}{\alpha_1(M_{\rm SUSY})}
\biggr]^{-\frac{11}{82}} ~, \nonumber \\[3pt]
 A_2 =&
A_L \cdot \biggl[
\frac{\alpha_3(M_{\text{SUSY}})}{\alpha_3(M_{\rm GUT})}
\biggr]^{\frac{4}{9}}
\biggl[
\frac{\alpha_2(M_{\text{SUSY}})}{\alpha_2(M_{\rm GUT})}
\biggr]^{-\frac{3}{2}}
\biggl[
\frac{\alpha_1(M_{\text{SUSY}})}{\alpha_1(M_{\rm GUT})}
\biggr]^{-\frac{23}{198}}
\nonumber \\[2pt]
&\times
\biggl[
\frac{\alpha_3(m_Z)}{\alpha_3(M_{\rm SUSY})}
\biggr]^{\frac{2}{7}}
\biggl[
\frac{\alpha_2(m_Z)}{\alpha_2(M_{\rm SUSY})}
\biggr]^{\frac{27}{38}}
\biggl[
\frac{\alpha_1(m_Z)}{\alpha_1(M_{\rm SUSY})}
\biggr]^{-\frac{23}{82}} ~.
\end{align}
Here, $A_L=1.25$ takes care of the long distance renormalization effects coming from QCD~\cite{Nihei:1994tx}.
The hadron matrix elements are calculated using lattice techniques in \cite{Aoki:2017puj} and are found to be

\begin{align}
\langle \pi^0\vert (ud)_Lu_R\vert p\rangle_1=\langle \pi^0\vert (ud)_Ru_L\vert p\rangle_1 &=
- 0.131(4)(13) ~\text{GeV}^2\ , \\
\langle \pi^0\vert (ud)_Lu_R\vert p\rangle_2=\langle \pi^0\vert (ud)_Ru_L\vert p\rangle_2 &=
- 0.118(3)(12) ~\text{GeV}^2
\ .
\end{align}

\subsection{Pure Gravity Mediation}
In order to calculate the proton lifetime in supersymmetric unification models, we need to specify the SUSY breaking spectrum. Although dimension-6 proton decay is not strongly dependent on the sparticle spectrum, the proton lifetime does depend on these masses through the running of the gauge couplings. The largest effect to the gauge couplings unification comes from incomplete representations of SU(5).
In the MSSM, the relevant particles are the gauginos and the Higgsinos.
In models like the constrained MSSM (CMSSM)~\cite{Drees:1992am,*Baer:1995nc,*Baer:1997ai,*Baer:2000jj,*Ellis:2001msa,Kane:1993td,*Ellis:1996xu,*Barger:1997kb,*Ellis:1997wva,*Ellis:1998jk,*Ellis:2000we,*Roszkowski:2001sb,*Djouadi:2001yk,*Ellis:2002rp,*Baer:2002gm,*Ellis:2003cw,*Baer:2003yh,*Lahanas:2003yz,*Chattopadhyay:2003xi,*Munoz:2003gx,*Arnowitt:2003vw,*Ellis:2010kf,*Ellis:2012aa,*Buchmueller:2013psa,Ellis:2012nv}, all supersymmetric particles are relatively degenerate. This means that the gauginos and the Higgsinos contribute to the running over roughly the same number of orders of magnitude. In contrast, models like pure gravity mediation have loop-suppressed gauginos,
due to their anomaly mediation origins~\cite{Randall:1998uk,Giudice:1998xp}%
\footnote{See Ref.\,\cite{Harigaya:2014sfa} for the path-integral derivation of the anomaly mediated gaugino mass.},
compared to the rest of the supersymmetry breaking spectrum.
This means that the gauginos contribute to the running of the gauge couplings over more energy scales than the rest of the SUSY particles. This type of SUSY spectrum flattens the running of $g_3$ and steepens the running of $g_2$ above the  gaugino masses,
which pushes up the GUT scale. This will have implications for the proton lifetime.
In the following discussion, we focus on this type of the SUSY spectrum,
that is a pure gravity mediation (PGM)~\cite{Ibe:2006de,Ibe:2011aa,Ibe:2012hu,Bhattacherjee:2012ed,ArkaniHamed:2012gw,Evans:2013lpa,Evans:2013dza,Evans:2014xpa,Evans:2014pxa} spectrum.
Product group models of unification tend to have a much too short proton lifetime. As we will see below, a PGM mass spectrum helps push up the unification scale which will have a mild but non-trivial effect on the proton lifetime.

Now, we examine the proton lifetime for the product group unification model found in~\cite{Ibe:2003ys}.  We use the SSARD code to evolve the gauge couplings, determine the supersymmetric spectrum, and calculate the proton lifetime~\cite{SSARD}. We take a PGM spectrum, which is described below, at the inputs scale defined where $g_1=g_2$. Since universal pure gravity mediation is rather restrictive, we will broaden our scope and included non-universal Higgs masses. This will make it easier to get the correct Higgs mass and allow us to see what kind of proton lifetimes DUNE and Hyper-K should expect. The code evolves the masses and couplings to the weak scale and checks that electroweak symmetry breaking is viable. The Higgs mass is also calculated to verify that it meets experimental constraints.

Before we present our results, we give a short review of pure gravity mediation. In pure gravity mediation, it is assumed that the field breaking supersymmetry is charged. Because of this charge, the gaugino masses are forbidden,  since the following operator is forbidden,
\begin{eqnarray}
W\supset \frac{c_g Z}{M_P}{\cal W W}~
\end{eqnarray}
where $\cal W$ is the gauge kinetic function, $Z$ is the SUSY breaking field, and $M_P$ is the Planck mass. The leading order contribution to the gaugino masses is then generated by anomaly mediation at the loop-level.
\begin{eqnarray}
M_i=\frac{b_ig_i^2}{16\pi^2} m_{3/2}~,
\end{eqnarray}
where $b_i={33/5,1,-3}$ for $i=1,2,3$.

The soft masses are a quite different. A charged supersymmetry breaking field, in general, cannot forbid the sfermion mass terms. The sfermions then get a mass of order $m_{3/2}$, from operators of the form
\begin{eqnarray}
K\supset \frac{c_{ij}|Z|^2}{M_P^2} \Phi_i^\dagger \Phi_j~.
\end{eqnarray}

The theory also has a supersymmetric bilinear mass.
\begin{eqnarray}
W\supset \mu_0 H_uH_d
\end{eqnarray}
where $H_{u,d}$ are the up and down Higgs masses respectively. Because the Higgs fields have an $R$-charge of zero in product group unificaiton models, the above Higgs bilinear term is forbidden. Even if the Higgs fields have zero R-charge, the Higgs bilinear mass terms can be generated in two ways. First, a Giudice-Masiero/Inoue-Kawasaki-Yamaguchi-Yanagida/Casas-Mu\~noz term~\cite{Giudice:1988yz,Inoue:1991rk,Casas:1992mk},
\begin{eqnarray}
\delta K = c_K H_uH_d+h.c.~,
\end{eqnarray}
will generate an effective Higgs bilinear mass of order $m_{3/2}$ as well a supersymmetric breaking $B$-term for the Higgs.
The other source of a Higgs bilinear term is from a non-renormalizable operator of the form
\begin{eqnarray}
\Delta W =\frac{c_W\langle W_h \rangle }{M_P^2}H_uH_d\ ,
\end{eqnarray}
where $W_h$ is the Hidden sector superpotential which has a non-zero vev and is responsible for generating the gravitino mass. Thus, this contribution to the Higgs bilinear mass is again of order $m_{3/2}$. If we consider both of these term, we find that the supersymmetric and supersymmetry breaking Higgs bilinear are linear independent,
\begin{align}
&\mu=(c_W +c_K)m_{3/2}\ ,\\
&B\mu=(-c_W+2c_K)m_{3/2}^2 \ ,
\end{align}
where $B$ is the supersymmetry breaking Higgs bilinear mass.

Since the simplest model found in \cite{Evans:2013lpa}, captures all the relevant features of PGM, we will start our examination with universal soft masses at the input scale and then extend our study to include non-universal Higgs masses. The free parameters of this theory are
\begin{eqnarray}
m_{3/2} \quad \quad \quad \tan\beta~.
\end{eqnarray}
The Higgs bilinear masses, $\mu$ and $B$, are determined by the electroweak breaking\footnote{This is equivalent to determining $c_{W,K}$.}, leaving just two free parameters. Since this will be rather restrictive, we further examine the case where the Higgs soft masses are free parameters giving the following set of free parameters
\begin{eqnarray}
m_{3/2} \quad \quad \quad \tan\beta \quad \quad \quad m_{H_u}^2 \quad \quad \quad m_{H_d}^2~.
\end{eqnarray}

\begin{figure}
\begin{center}
\begin{minipage}{8in}
\includegraphics[trim={0.7in 0 .7in 0},height=2.75in]{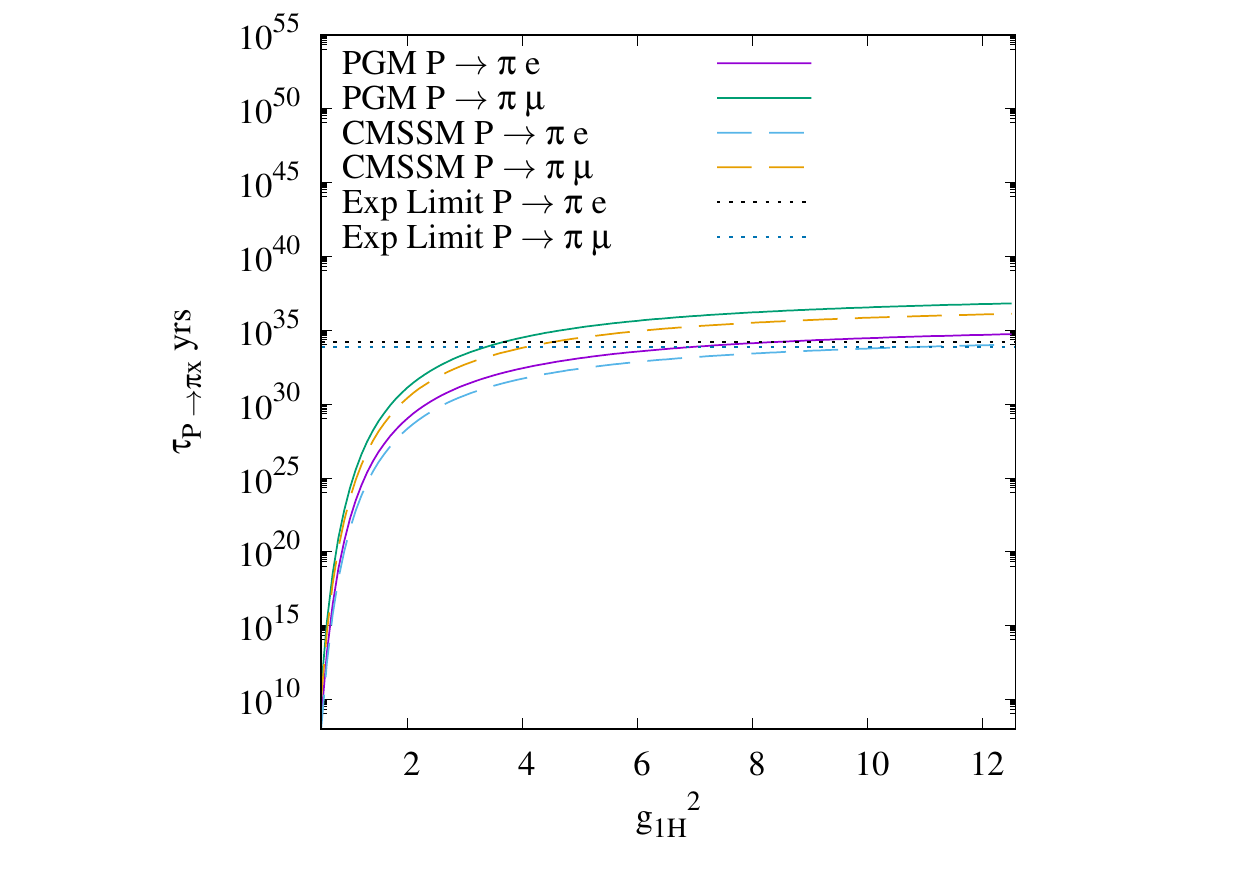}
\hspace*{.25in}
\includegraphics[trim={1in 0 0.7in 0},height=2.75in]{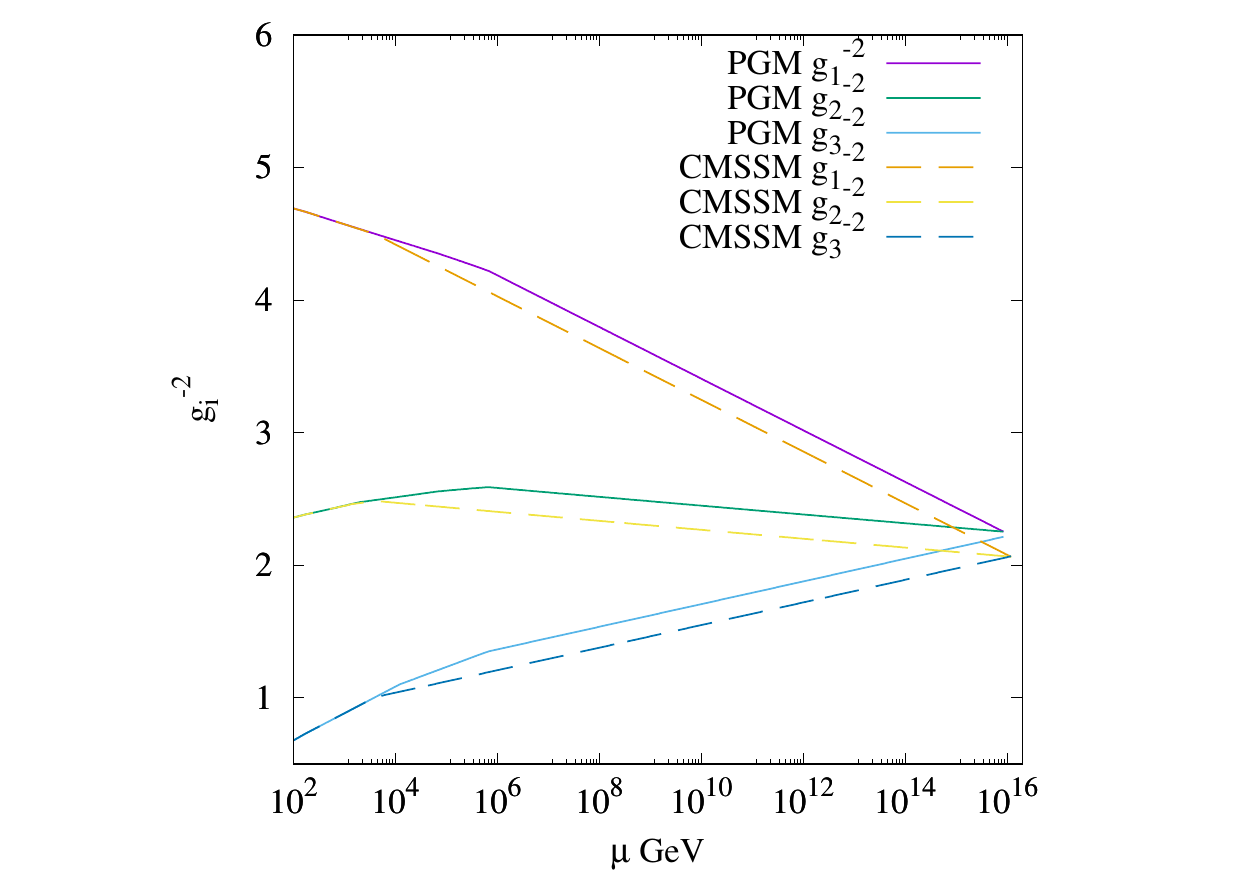}
\end{minipage}
\caption{\label{fig:gaugeCom}
Left) The proton lifetime as a function of
$g_{1H}^2$ at the GUT scale of $\order{\langle Q\rangle}$.
The perturbative GUT below the cutoff scale $\Lambda = 4\pi \langle Q\rangle$ is achieved for $g_{1H}^2 \le 3.68$ (see Eq.\,\eqref{eq:g1hlandau}).
The horizontal dotted lines show the
current experimental limits on the proton life time of the modes, $p\to \pi^0+e^+$
and $p\to\pi^0+\mu^+$~\cite{Miura:2016krn}, respectively.
Right) The coupling unification for given SUSY spectrum.
The better the couplings unify,
the lower $M_X$ is for a given $g_{1H}^2(\langle Q\rangle)$ (see Eq.\,\eqref{eq:MXg1h}).
}
\end{center}
\end{figure}

\subsection{Proton Lifetime}
In this section, we present the results of our calculation of the proton lifetime for minimal product group unification discussed above. In Fig.\,\ref{fig:gaugeCom}, we compare the $g_{1H}^2$ dependence of the proton lifetime for a pure gravity mediation spectrum with $m_{3/2}=700$ TeV, $\mu<0$ and $\tan\beta=2.1$ to that for a CMSSM spectrum with $m_{1/2}=1.75$ TeV $m_0=4$ TeV $A_0/m_0=2$, $\mu>0$, and $\tan\beta=20$. These values are chosen to obtain a relatively good Higgs mass, however, varying these numbers will not change our conclusions significantly. As is clearly seen, the lifetime is smaller by a non-trivial amount for the CMSSM spectrum. This is ultimately due to the fact that the gauge couplings unify better in the CMSSM, which is also seen in Fig.\,\ref{fig:gaugeCom}.

Fig.\,\ref{fig:gaugeCom} also shows the sharp suppression of the proton lifetime near $g_{1H}^2=1$. This is due to the exponential suppression of $M_X$ as $g_{1H}^2$ becomes smaller at the GUT scale, see Eq.\,\eqref{eq:MXg1h}.
We see that the proton lifetime is too short unless $g_{1H}^2\gtrsim 6$ even for a PGM spectrum.
This contradicts the constraint coming from well defined Yukawa couplings which requires $g_{1H}^2\lesssim 3.68$ as seen in Eq.\,\eqref{eq:g1hlandau}. Thus, this minimal model of product group unification is ruled out. This leads us to consider non-minimal models of product group unification.

\section{Light Colored Particles}
\subsection{SU(5) incomplete multiplet below the GUT scale}
As we saw in the previous section, proton decay constraints rule out the simplest product group unification models. The short lifetime of the proton is attributed to the light $X$ bosons, which was a consequence of our separations of the cutoff scale and the GUT scale.

A simple way to address the proton lifetime problem is to add additional representations of SU(5) with the SU(2) and SU(3) components having different masses.
In fact, product unification model
can easily achieve the SU(5) incomplete multiplets by
introducing an additional $\mathbf{5},\mathbf{\bar 5}$ ($\Phi',\bar \Phi'$ in Table\,\ref{table1}) and $2,\bar 2$ ($\Theta,\bar\Theta$ in Table\,\ref{table1}).
These fields are coupled to the SU(5)$\times$U(2)$_{\mathrm{H}}
$ breaking fields, $Q,\bar Q$, in the following way,
\begin{eqnarray}
\Delta W = \lambda \Phi'Q \bar \Theta +\lambda \bar \Phi' \bar Q \Theta +\mu_5 \Phi'\bar \Phi'~.
\end{eqnarray}
Here, we have taken the Yukawa couplings equal for simplicity. In this model, the doublets of the $\Phi', \bar{\Phi}'$ pair up with the $\Theta,\bar \Theta$ and obtain mass from the VEV of $Q,\bar Q$, while the triplets of $\Phi', \bar{\Phi}'$ obtain the mass of $\mu_5$.
In this way, the SU(5) incomplete multiplets below the GUT scale are achieved without fine-tuning.

If $\mu_5\ll \lambda \langle Q\rangle$, the matching conditions in Eqs.\,\eqref{eq:g3mat}--\eqref{eq:g1mat} get non-trivial corrections. The expressions for $M_X$ is then modified to
\begin{eqnarray}
M_X\simeq\left(\frac{M_{G_0}}{\mu_5}\right)^{\frac{2}{15}}M_{X_0} \label{eq:MXg1hDel}
\end{eqnarray}
where $M_{G_0},M_{X_0}$ are the scale the coupling unify at and the heavy gauge boson mass for the case without the additional $\Phi',\bar \Phi'$ respectively.
We have taken $\lambda v=M_G$ to maximize the effect of $\Phi',\bar \Phi'$. Because $M_X$ scales vary slowly with $\mu_5$ in this expression, we will have to take $\mu_5\ll M_G$.

Now, we look at the modifications to the RG running of the hidden sector gauge couplings from these additional states. The beta function of $g_{1H}$ is modified to \begin{eqnarray}
\frac{d g_{1H}^2}{d\ln\mu}= 7\frac{g_{1H}^4}{8\pi^2}
\end{eqnarray}
giving
\begin{eqnarray}
g_{1H}^2\left({\Lambda}/{4\pi}\right)=3.29\label{eq:g1hlandau}
\end{eqnarray}
if $g_{1H}^2\left(\Lambda\right)=4\pi$. This slight modification to the upper bound on $g_{1H}^2$
can easily be offset by reducing $\mu_5$. As we will see below, this will allow us to get a proton lifetime beyond the current experimental limit.  However, because of the slow scaling of $M_X$ seen in Eq.\,\eqref{eq:MXg1hDel}, the proton lifetime still has an upper limit.

\subsection{Suppressing
\texorpdfstring{$\mu_5$}{Lg}
 with a PQ-Symmetry}
Before we present our results, we wish to motivate the suppression of $\mu_5$ below the GUT scale.
The most attractive possibility is
the Peccei-Quinn (PQ) symmetry
which solves the Strong CP problem~\cite{Peccei:1977hh,Peccei:1977ur}.
In the PQ mechanism,
$\theta$-angle of QCD is
canceled by the VEV of the
axion~\cite{Weinberg:1977ma,Wilczek:1977pj}  associated with the spontaneous breaking of the PQ symmetry.

The PQ mechanism is particularly attractive when its breaking scale is of $10^{10\mbox{--}12}$\,GeV for which the axion is a natural candidate for cold dark matter.
For example, if the PQ-breaking is broken before inflation, the axion dark matter density is given by the misalignment mechanism,
\begin{eqnarray}
\Omega_ah^2= 0.18\theta_a^2\left(\frac{F_a}{10^{12}~{\mathrm{GeV}}}\right)^{1.18}\left(\frac{\Lambda_\mathrm{QCD}}{400~{\mathrm{MeV}}}\right)\ .
\end{eqnarray}
Here, $\theta_a$ is the initial misalignment angle of the axion, $F_a$ is the PQ breaking scale, and $\Lambda_\mathrm{QCD}$ is the QCD scale~\cite{Turner:1985si} (see also \cite{Borsanyi:2016ksw,Ballesteros:2016xej}).%
\footnote{In this case,  the quantum fluctuation of the axion during inflation leads to the isocurvature fluctuation of the axion dark matter density.
As its amplitude is proportional to the Hubble parameter during inflation, $H_I$,
the cosmic microwave background constraints on the isocurvature fluctuation puts a constraint,
$H_I\lesssim 10^{7-8}$\,GeV, when the axion is the dominant dark matter~(see e.g. \cite{Kawasaki:2013ae}.)}
If the PQ-breaking takes place after inflation, on the other hand,
cosmic strings are formed at the phase transition of the PQ breaking.
The axion winds $N_\mathrm{w}(\ge 1)$-times around the cosmic string, and hence, the cosmic string is attached by $N_\mathrm{w}$ domain walls when the axion obtains a non-trivial scalar potential due to the QCD effect.
For a model with $N_\mathrm{w}>1$, the string-wall network is stable and dominates the energy density immediately, which is not consistent with our Universe.
For a model with $N_\mathrm{w}=1$,
the string-wall network is unstable, and it disappears immediately.
In this case, the axion dark matter is dominated by the contributions emitted from the decay of the string-wall network~\cite{Hiramatsu:2012gg} (see also \cite{Ringwald:2018xlf}),
\begin{align}
    \Omega_ah^2 = 0.035 \pm 0.012
    \left(
    \frac{F_a}{10^{10}\,\mathrm{GeV}}
    \right)^{1.19}\left(\frac{\Lambda_\mathrm{QCD}}{400~{\mathrm{MeV}}}\right)\ .
\end{align}
The PQ-breaking scale is also constrained to be $F_a \gtrsim 10^{9}$\,GeV from astrophysical phenomena~\cite{Raffelt:2006cw,Chang:2018rso,Irastorza:2018dyq,Hamaguchi:2018oqw}.
From these considerations,
we assume the PQ-breaking scale of $10^{10\mbox{--}12}$\,GeV in the following discussion.

To associate $\mu_5$ with the PQ-symmetry breaking,
we introduce a PQ symmetry breaking field $P$, in which the axion resides as,
\begin{align}
P = \frac{F_a}{\sqrt{2}} e^{-i a / F_a}\ .
\end{align}
The PQ charges can be found in  Table\,\ref{table1}.%
\footnote{We set the PQ-charge of $\bar{\Phi}'$ vanishing, which allows slight mixing between $\bar\Phi$ and $\bar\Phi'$ through which the triplets in $\Phi'$ and $\bar \Phi'$ can decay into the MSSM fields.
A sizable mixing between $\bar\Phi$ and $\bar\Phi'$ affects
the proton lifetime and the branching fractions~\cite{Ibe:2019ifm}.
}
As there is only one pair of $\Phi'$, $\bar{\Phi}'$, the domain wall number $N_\mathrm{w}=1$ in this model.
With this PQ charge assignment, the supersymmetric mass term for
$\Theta\bar \Theta$ is forbidden as is the $\mu_5$ term and we instead have
\begin{eqnarray}
\Delta W = \lambda \Phi'Q \bar \Theta +\lambda \bar \Phi' \bar Q \Theta +\lambda_P P\Phi'\bar \Phi'~,
\end{eqnarray}
with $\lambda$'s being the coupling constants of order of unity.
The effective $\mu_5$ is then given by $\mu_5  = \lambda_P \langle P \rangle $.
The VEV of $Q,\bar Q$ give masses only to the doublets, while
the triplets of $\Phi', \bar\Phi'$
obtain the much smaller mass,
$\mu_5\ll \lambda_P \langle Q\rangle = \order{M_G}$.

As we will see shortly, this connection of $\mu_5$ and the PQ breaking scale will put the proton lifetime in reach of coming proton decay experiments
due to Eq.\,\eqref{eq:MXg1hDel}. Furthermore, the possibility of axion dark matter make the parameter space of models like pure gravity mediation much less restricted, as we will see below.

Before closing this section, let us comment on the axion coupling to photons through the the electromagnetic anomaly,
\begin{align}
    {\cal L} = \frac{1}{4}G_{a\gamma\gamma}a F\tilde{F}\ ,
\end{align}
where $F$ and $\tilde{F}$ denote the QED field strength and its dual.
The coupling constant $G_{a\gamma\gamma}$ is given by,
\begin{align}
G_{a\gamma\gamma} =
\frac{\alpha}{2\pi}\left(c_{a\gamma\gamma} - \frac{2}{3}\frac{1+4z}{1+z}\right)
\frac{1+z}{z^{1/2}} \frac{m_\pi}{m_a}\frac{1}{f_\pi}\ ,
\end{align}
where $f_\pi \simeq 92$\,MeV, $z= m_u/m_d \simeq 0.553\pm0.043$~\cite{Leutwyler:1996qg}. Here we have inserted the axion mass,
\begin{align}
m_a = \frac{z^{1/2}}{1+z} \frac{f_\pi }{F_a}m_\pi \ .
\end{align}
In the present model, $c_{a\gamma\gamma}$ is given by $c_{a\gamma\gamma}= 2/3$, which should be compared with the complete GUT KSVZ multiplet of $\mathbf{5}$, $\mathbf{\bar{5}}$ giving $c_{a\gamma\gamma} = 8/3$~(see e.g.~\cite{diCortona:2015ldu}).
As a result, the present model predicts an axion coupling to QED which is three times larger, for a given axion mass, than in the conventional GUT model with a complete KSVZ multiplet.

\section{Results}
\begin{figure}[t!]
\begin{center}
\includegraphics[trim={1in 0 1.25in 0},width=.5\textwidth]{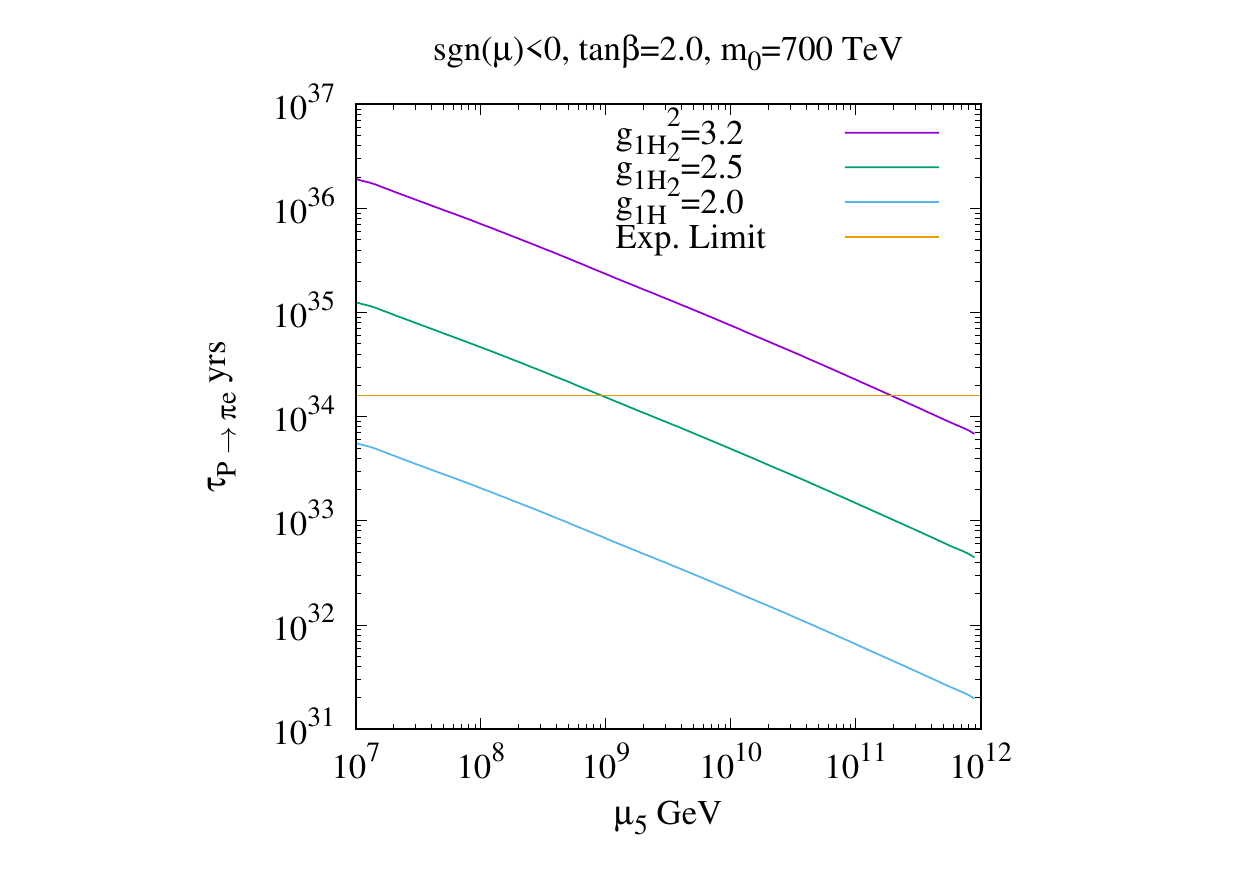}
\caption{\label{fig:muvtau}$\mu_5$ dependence of proton lifetime, $p\to \pi^0+e^+$ in PGM for given value of $g_{1H}^2$ at the GUT scale.}
\end{center}
\end{figure}
Here we show the results of our study of product group unification with the addition of a pair of light triplet quarks. We begin with universal pure gravity mediation and study the effect of $\mu_5$ on the lifetime of the proton.
In Fig.\,\ref{fig:muvtau}, we plot the proton lifetime versus $\mu_5$ for different values of $g_{1H}^2$ for a PGM spectrum with $m_{3/2}=700$\,TeV, $\tan\beta=2$, and $\mu>0$. As can be seen in this figure, the largest lifetime occurs for larger values of $g_{1H}^2$ and smaller values of $\mu_5$.  However, the separation of strong dynamics from the GUT scale limits how large we can take $g_{1H}^2$, Eq.\,\eqref{eq:g1hlandau}. Because of our naive estimation for $g_{1H}^2$ in Eq.\,\eqref{eq:g1hlandau}, we take $g_{H_1}^2=3.2$ as the maximal value of $g_{1H}^2$ in Fig.\,\ref{fig:muvtau} using it only as a guide. However, as can be estimated from the figure, a small change in $g_{1H}^2$ does not affect our conclusions too much. Furthermore, from Fig.\,\ref{fig:muvtau}, it is clear that the lifetime scales quite close to our estimate in Eq.\,\eqref{eq:MXg1hDel} and it is quite difficult to push $\mu_5$ beyond $10^{12}$\,GeV.

\begin{figure}[!t]
\begin{minipage}{8in}
\includegraphics[trim={0.7in 0 .7in 0},height=3.in]{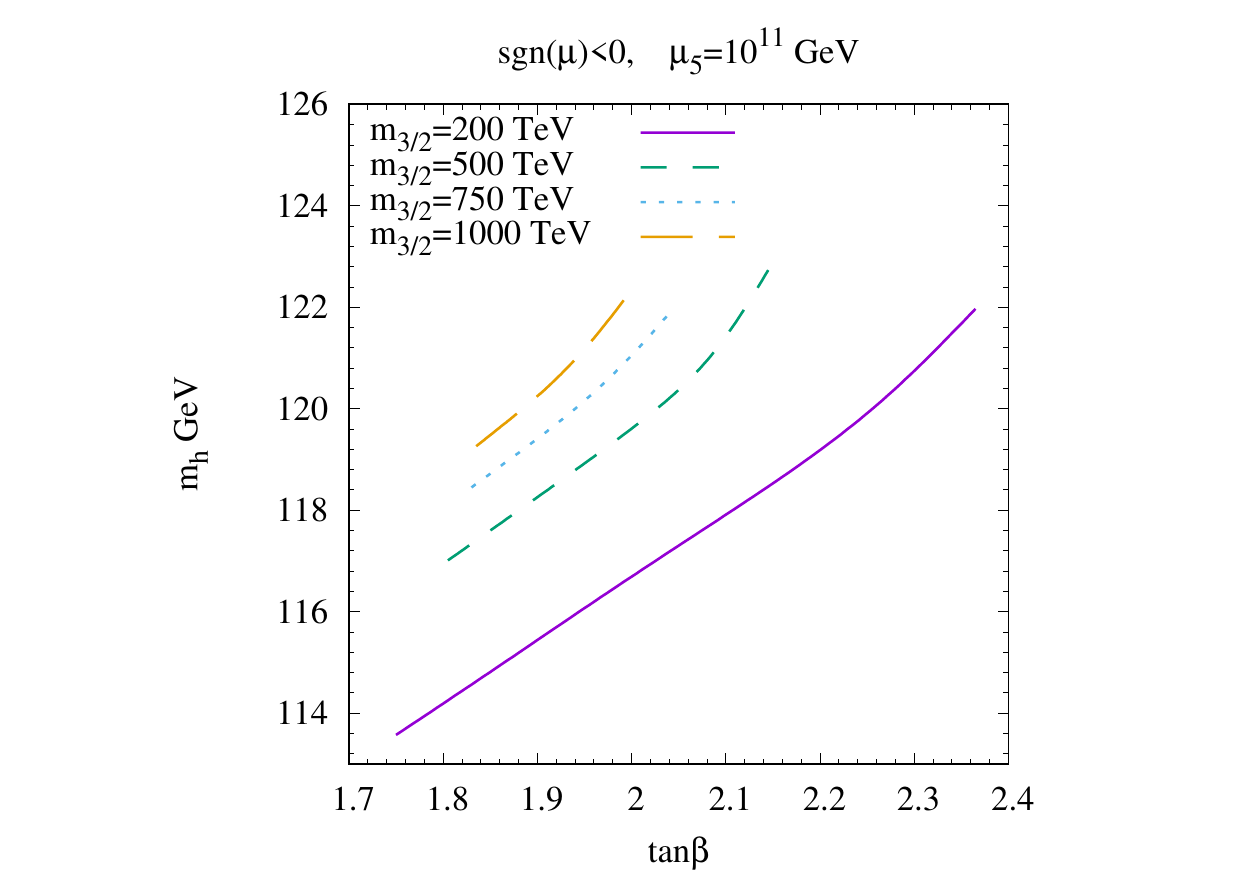}
\includegraphics[trim={0.7in 0 0.7in 0},height=3.in]{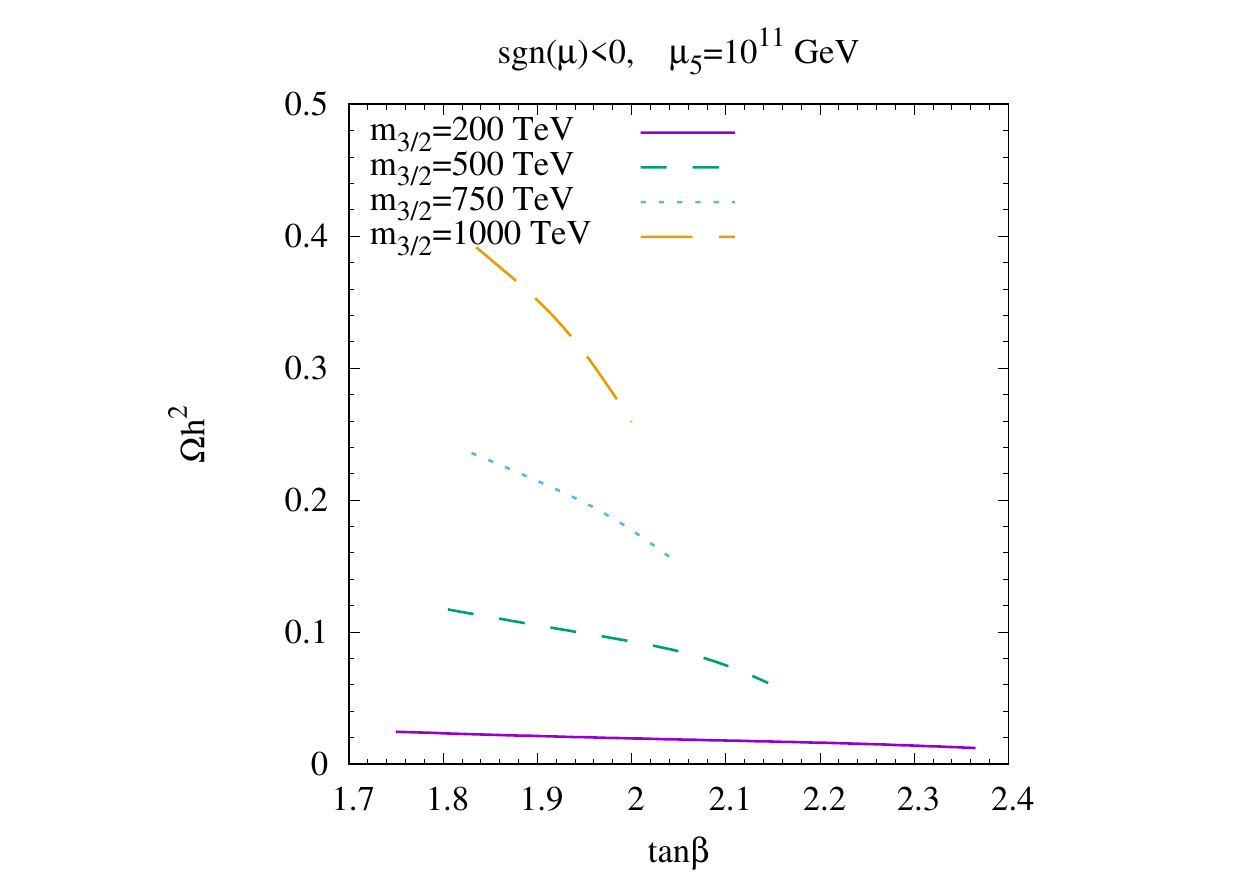}
\end{minipage}

\begin{minipage}{8in}
\begin{center}
\hspace{-1.5in}\includegraphics[trim={0.9in 0 0.7in 0},height=3.in]{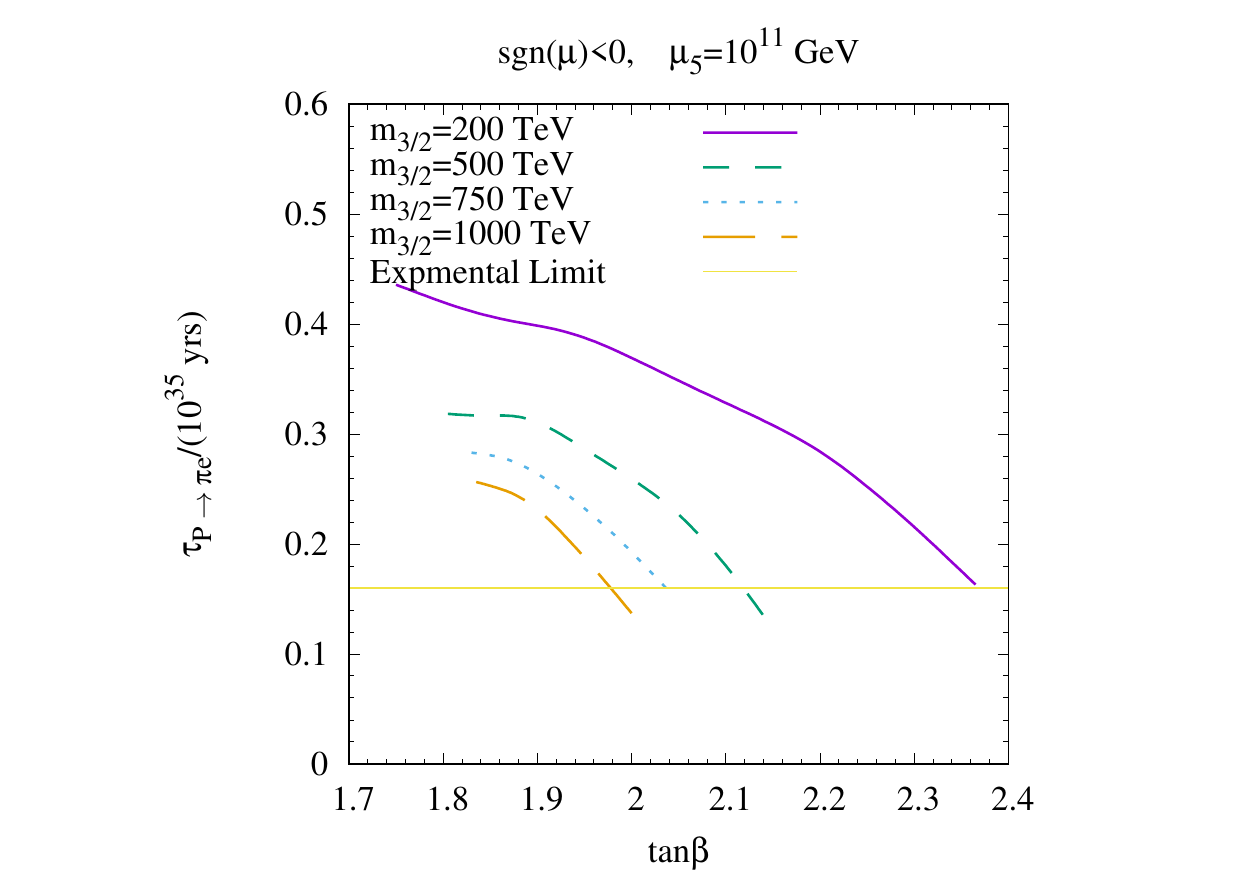}
\end{center}
\end{minipage}

\caption{\label{fig:UPGM}
The lightest Higgs boson mass, $m_h$, the LSP abundance, and the proton lifetime for the universal PGM as a function of $\tan\beta$.
}

\end{figure}

\begin{figure}[!t]
\begin{minipage}{8in}
\includegraphics[trim={0.7in 0 .7in 0},height=3.in]{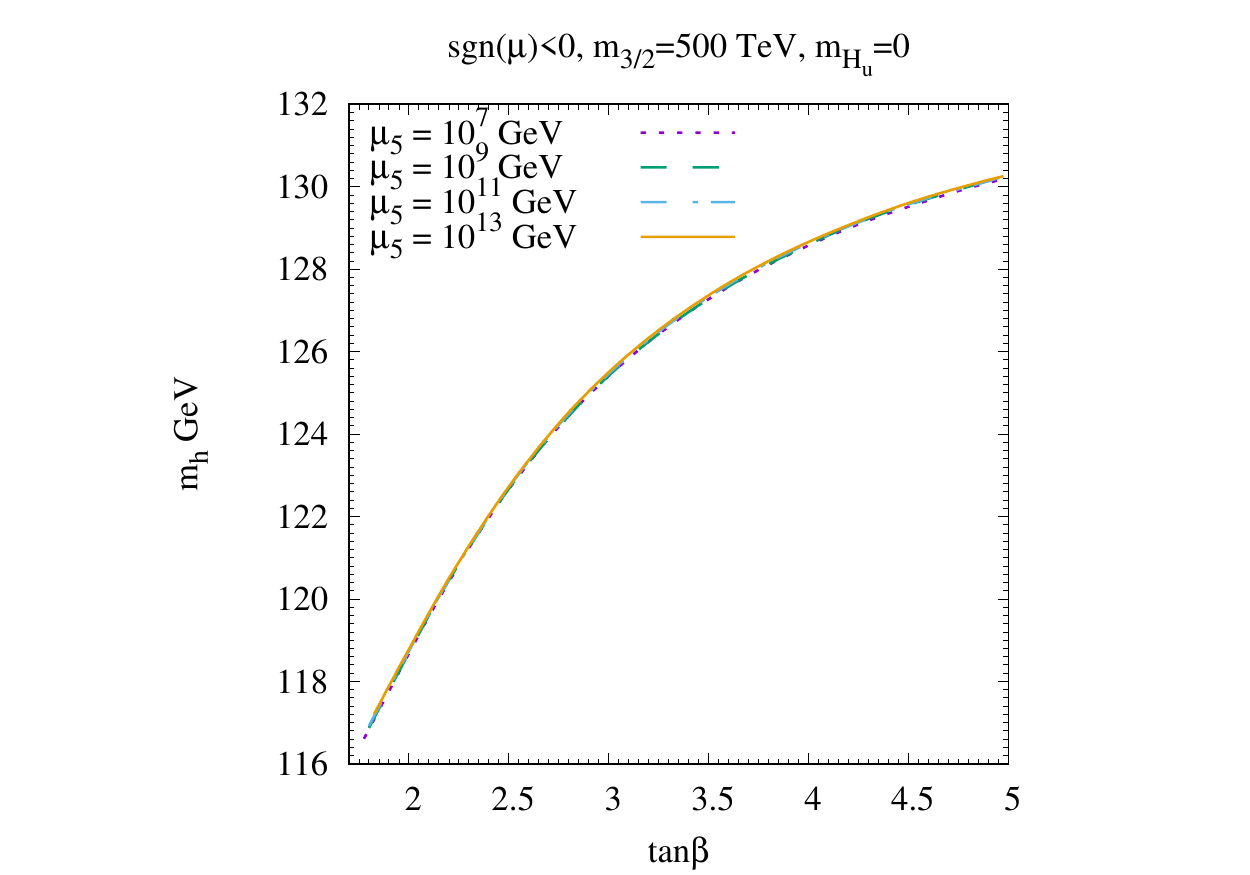}
\includegraphics[trim={0.7in 0 0.7in 0},height=3.in]{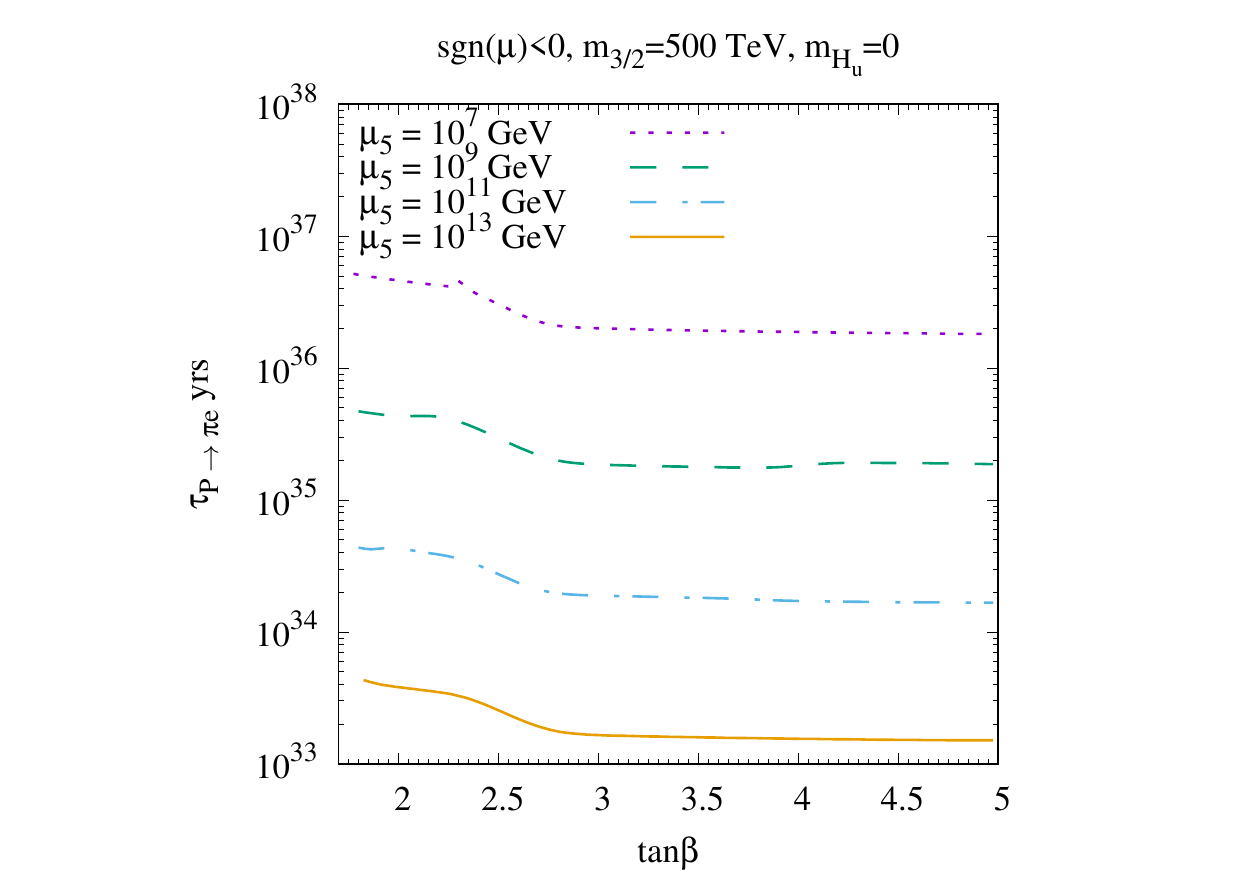}
\end{minipage}

\begin{minipage}{8in}
\includegraphics[trim={0.7in 0 .7in 0},height=3.in]{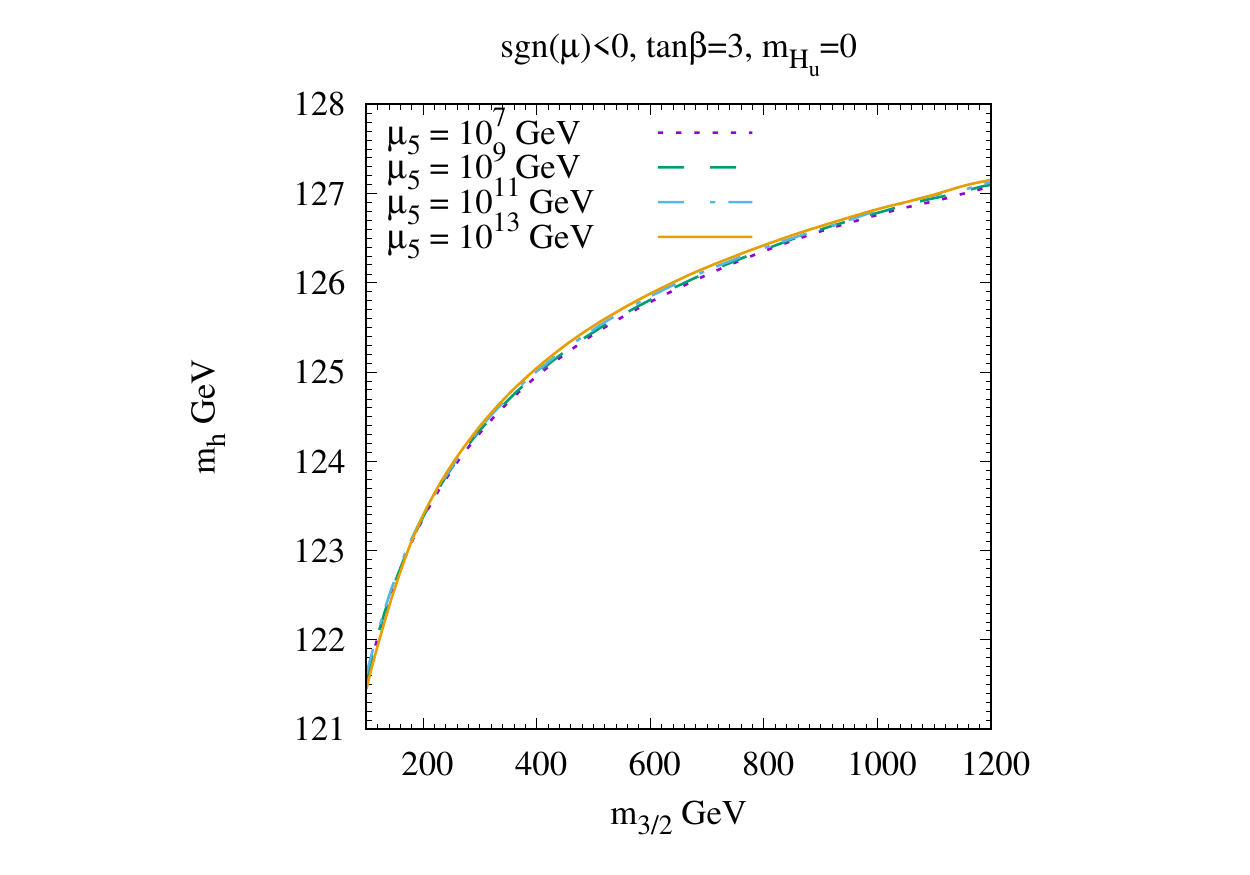}
\includegraphics[trim={0.7in 0 0.7in 0},height=3.in]{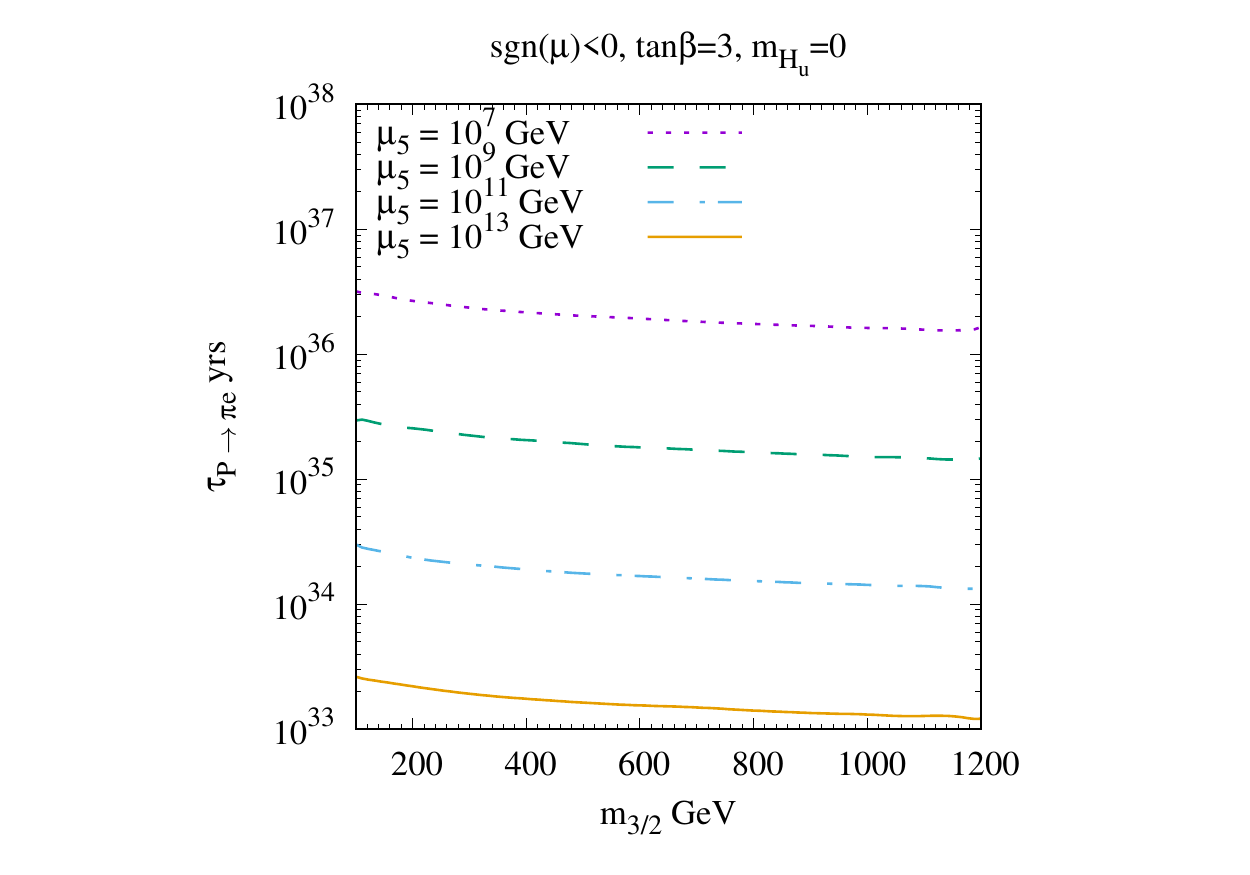}
\end{minipage}

\caption{\label{fig:mhu0mu5}
The $\tan\beta$ (top) and the $m_{3/2}$ (bottom) dependence of $m_h$ (left) and of the proton lifetime (right). Here, we take $m_{H_u}^2 = m_{3/2}^2$ and $m_{H_d}^2 = 0$ as an example of the non-universal Higgs mass.
}

\end{figure}

Next, we consider universal PGM with the addition of $\Phi',\bar \Phi'$. Our results can be seen in Fig.\,\ref{fig:UPGM}. The lines terminate at larger $\tan\beta$ due to electroweak symmetry breaking (EWSB) failing, i.e. equations need $|\mu|^2<0$. For smaller $\tan\beta$, the lines terminate due to a non-perturbative Yukawa coupling. Although the edge with larger $\tan\beta$ may be allowed due to the large errors in calculating the Higgs mass, much of the parameter space is still ruled out by the Higgs mass measurement, $m_h = 125.10\pm0.14$\,GeV~\cite{Zyla:2020zbs}.
With $\mu_5=10^{11}$\,GeV, the proton lifetime is sufficiently long for most of the parameter space.  However, for the edge with larger $\tan\beta$, where the Higgs mass is most consistent with the measured value, the proton lifetime tends to be too short. This means universal PGM needs $\mu_5< 10^{11}$\,GeV. In the figure, we also show
the thermal relic lightest supersymmetric particle (LSP)
contribution to the dark matter abundance.
The LSP is mostly Wino-like neutralino in the PGM spectrum. For $m_{3/2}\lesssim 500$ TeV, the Wino dark matter density is insufficient to explain the measured value. However, for this range of $m_{3/2}$ and value of $\mu_5$, the axion can make up a large fraction of the dark matter.

Since it is very possible that the soft masses of PGM are non-universal, we will look at more generic mass spectra.  Since the dimension-6 proton decay is quite insensitive to the sfermion masses, we will consider the case where only the Higgs soft masses are non-universal.
This will relax the tension on the proton lifetime coming from the Higgs mass measurement and allow us to more fully explore the parameter space consistent with product group unification.

We begin our study of non-universal Higgs masses by looking at the dependence of the Higgs boson mass and proton lifetime on $m_{3/2}$ and $\tan\beta$ for different values of $\mu_5$.
To demonstrate the non-universal Higgs masses, we take $m_{H_d}^2 = m_{3/2}^2$ while $m_{H_u}^2=0$.
With the non-universal Higgs mass,
the successful EWSB is achieved even for $\tan\beta \gtrsim 2$.
As is clear from Fig.\,\ref{fig:mhu0mu5}, the lightest Higgs mass strongly depends on $\tan\beta$ and $m_{3/2}$ but vary mildly with $\mu_5$.
This is because the only effect of $\mu_5$ is to mildly change the running of the gauge couplings.
However, since the couplings are fixed by experiment at the low-scale, this effect is quite mild. The proton lifetime, on the other hand, depends quite mildly on both $\tan\beta$ and $m_{3/2}$ and very strongly on $\mu_5$. This is due to the fact that $\mu_5$ can have a significant effect on the running of the gauge couplings for scales above $\mu_5$. $\tan\beta$ and $m_{3/2}$, in contrast, only affect the gauge coupling running indirectly through the Higgsino's and gaugino's masses.

The effects discussed above can be see in Fig.\,\ref{fig:mhu0mu5}. In the top two figures, we see the typical strong dependence of the lightest Higgs mass dependence on $\tan\beta$. The current experimental limit, $m_h = 125.10\pm0.14$\,GeV~\cite{Zyla:2020zbs}, combined with the theoretical uncertainties, constrains $\tan\beta$ to be roughly in the range $2.5\mbox{--}4$.
The proton lifetime is, as expected, quite mildly dependent on $\tan\beta$ and saturates at about $\tan\beta\simeq 3$. This is due to a saturation of the Higgsino and Wino masses $\tan\beta$ dependence.%
\footnote{The Higgsino's $\tan\beta$ dependence is through the electroweak symmetry breaking conditions, which are how the Higgsino mass, $\mu$, is determined. The Wino gets a $\tan\beta$ dependence through a relatively large threshold correction generated when the Higgsino and heavy Higgs bosons are integrated out.}


\begin{figure}[t!]
\begin{center}
\includegraphics[trim={1in 0 1.25in 0},width=.4\textwidth]{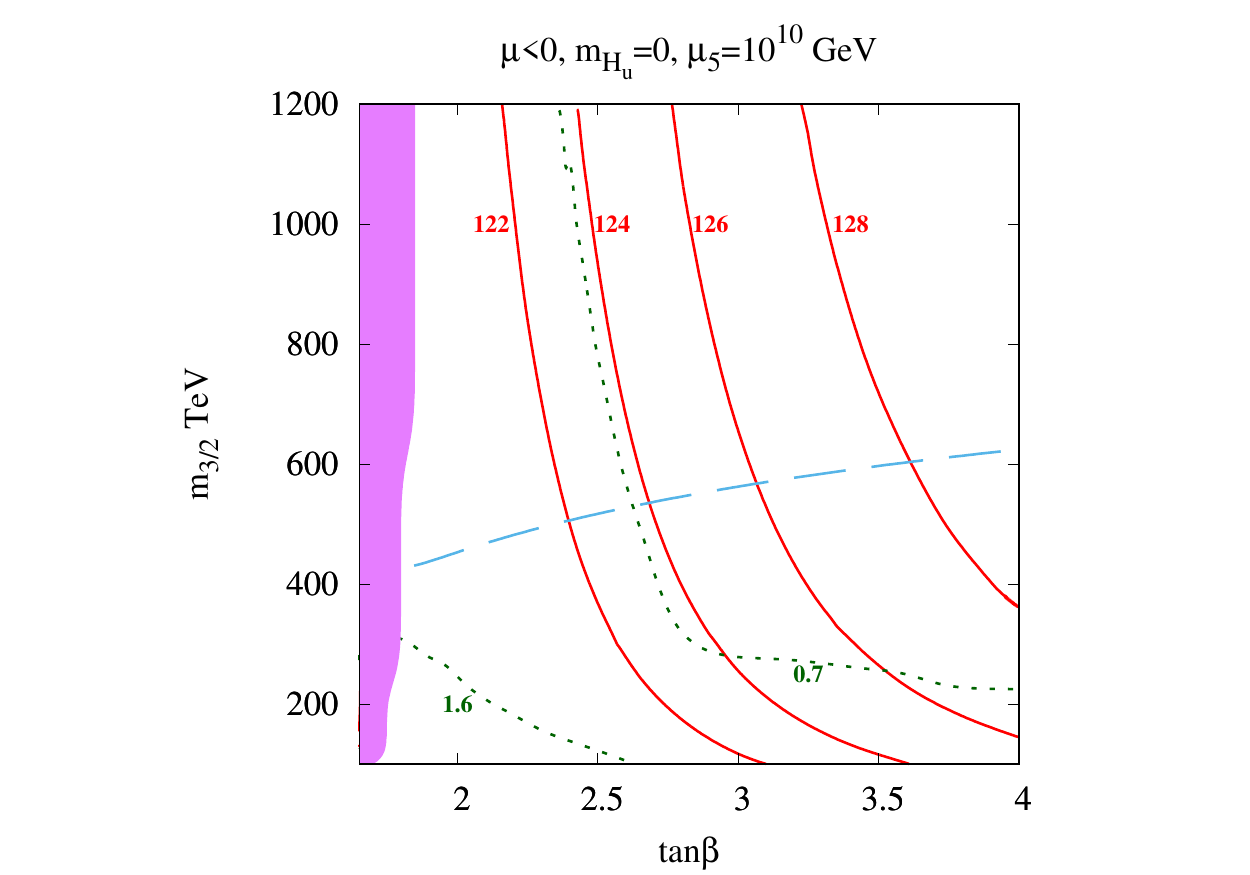}
\caption{\label{fig:tbvsm32mhu0}
The $\tan\beta$ and $m_{3/2}$ dependence of the lightest Higgs boson mass in units of GeV (red lines) and the proton lifetime in units of $10^{35}$\,years (green dotted lines).
Here, we take $m_{H_u}^2 = 0$ and $m_{H_d}^2 = m_{3/2}^2$ as an example of the non-universal Higgs mass.
The blue dashed line is where the Wino masses give the correct thermal relic density for dark matter.
The pink shaded region is excluded by the
non-perturbative Yukawa coupling.
}
\end{center}
\end{figure}

For the bottom two figures in Fig.\,\ref{fig:mhu0mu5}, we see the expected strong dependence of the lightest Higgs boson mass on $m_{3/2}$, which determines all the sfermion masses.
Due to the theoretical uncertainties in Higgs mass calculation, all plotted values are consistent with the measure Higgs boson mass. The proton lifetimes dependence on $m_{3/2}$ is through the Higgsino and Wino. The Wino mass is generated through anomaly mediation with a non-trivial threshold correction coming from the Higgsinos and Heavy Higgs. Since, the Higgsino mass also scales with $m_{3/2}$, it is roughly set by the stop mass. However, since this mass dependence of the gauge couplings is only logarithmic, we only see a mild dependence of the proton lifetime on $m_{3/2}$ in Fig.\,\ref{fig:mhu0mu5}.

 \begin{figure}[!t]
\begin{minipage}{8in}
\includegraphics[trim={0.7in 0 .7in 0},height=3.in]{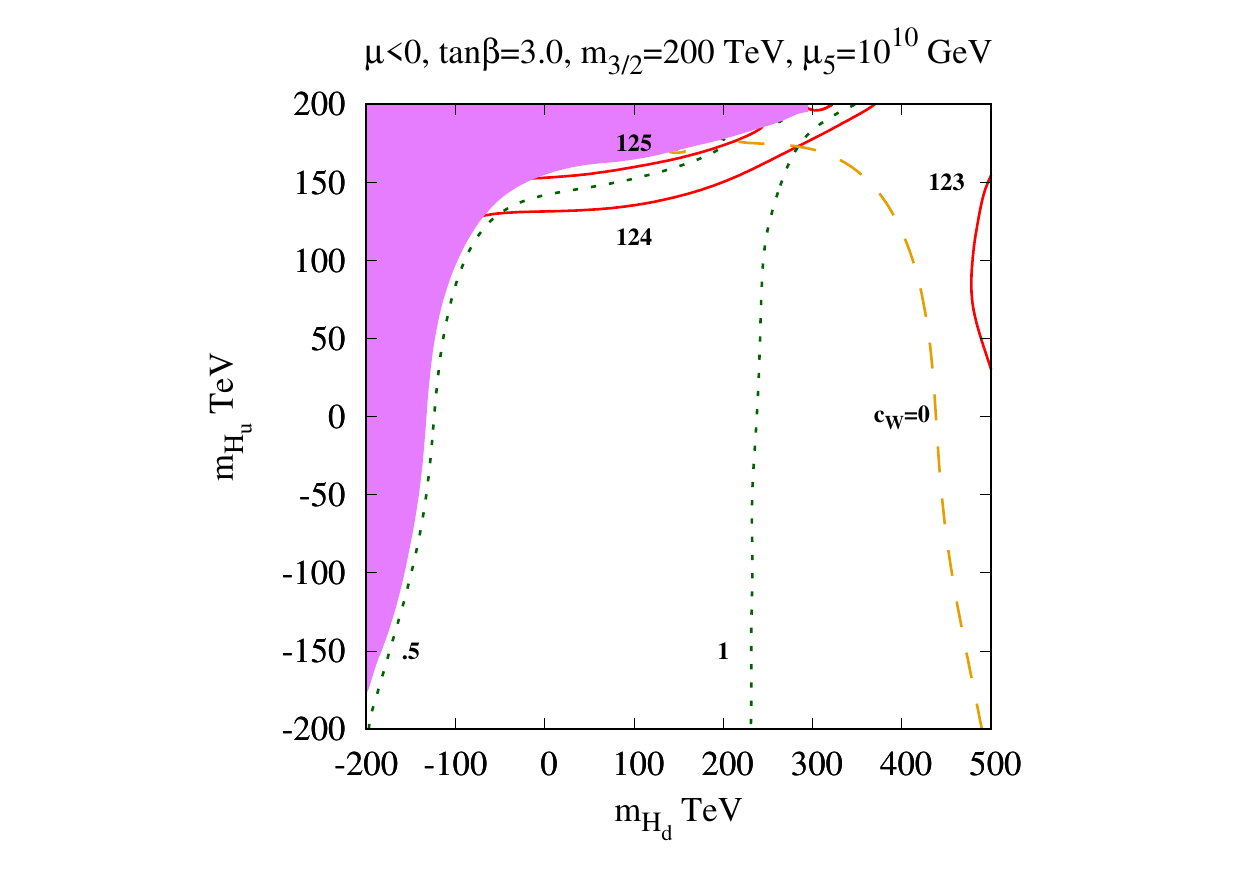}
\includegraphics[trim={0.7in 0 0.7in 0},height=3.in]{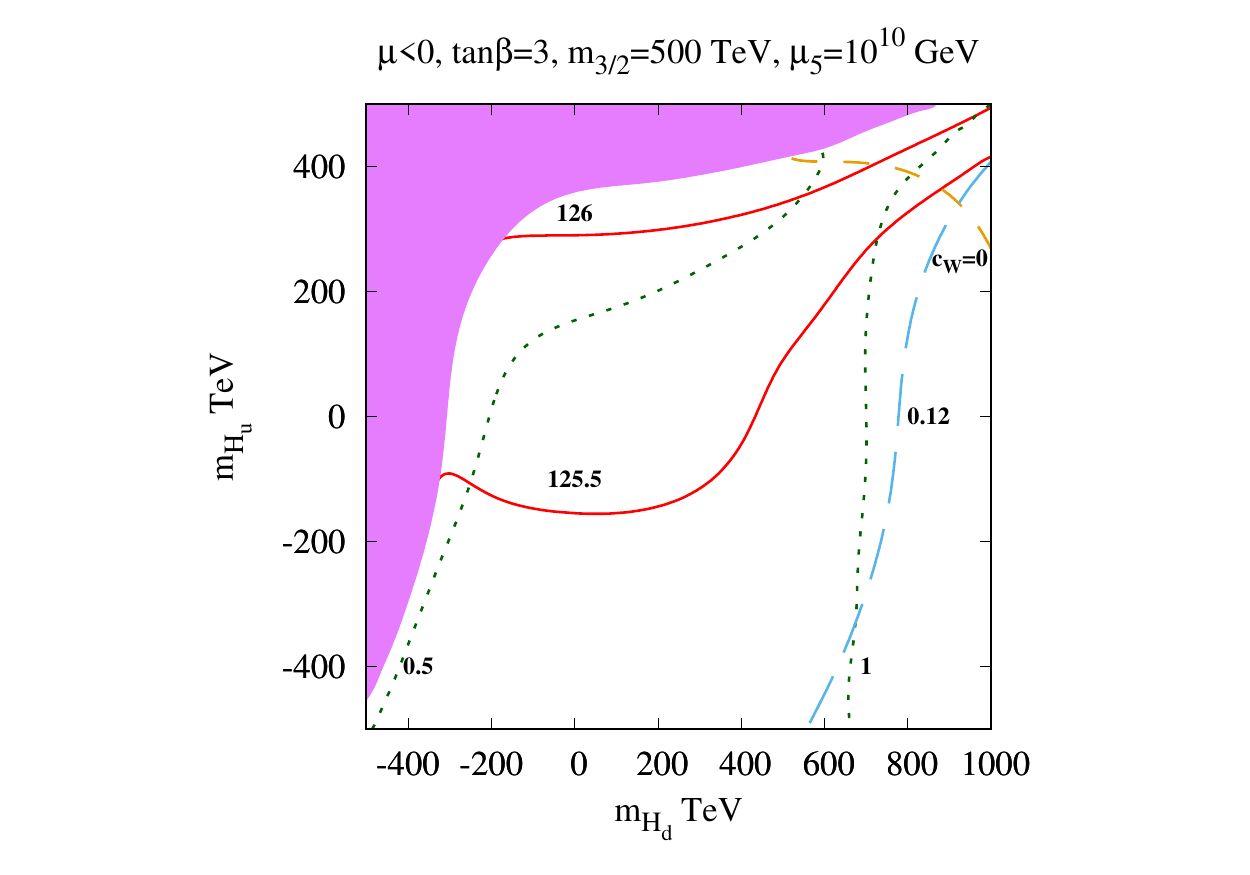}
\end{minipage}

\begin{minipage}{8in}
\includegraphics[trim={0.7in 0 .7in 0},height=3.in]{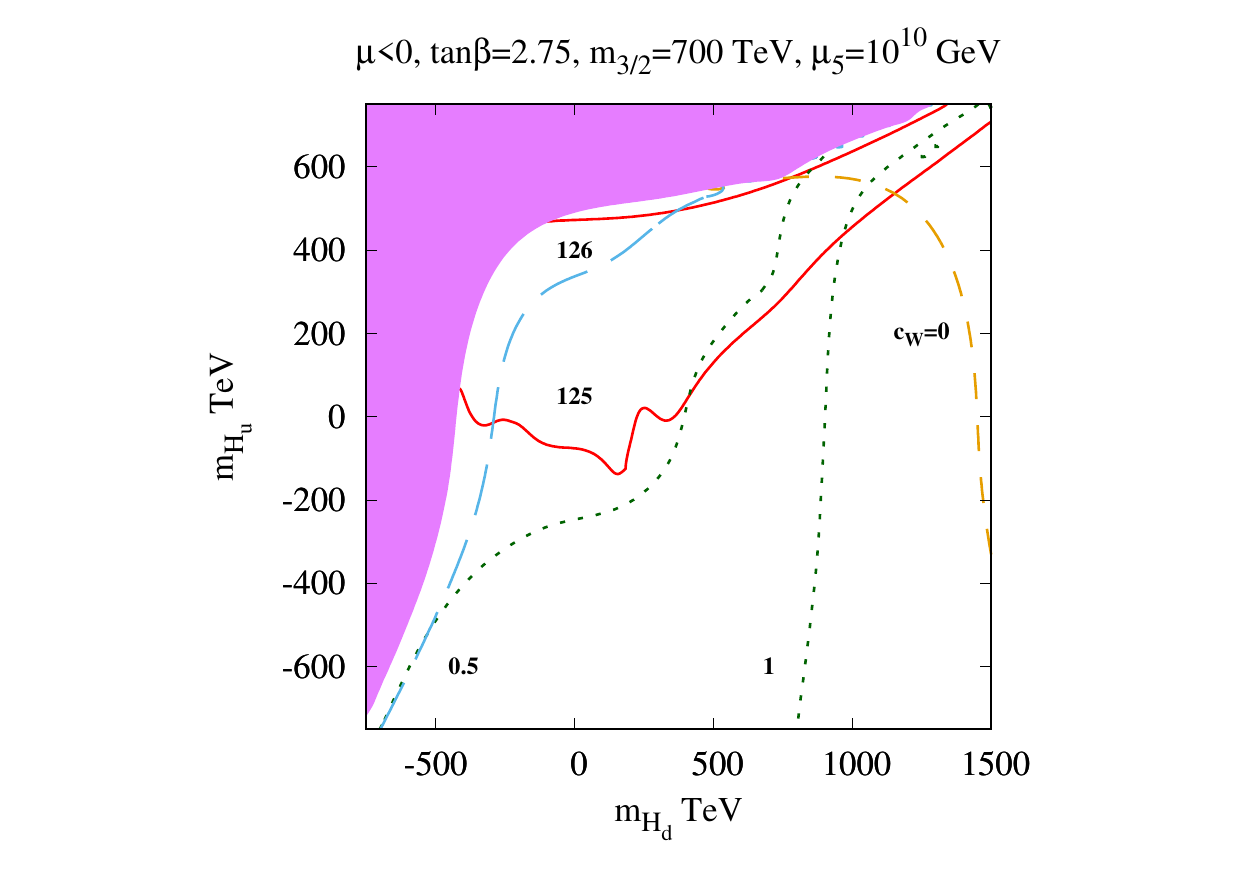}
\includegraphics[trim={0.7in 0 0.7in 0},height=3.in]{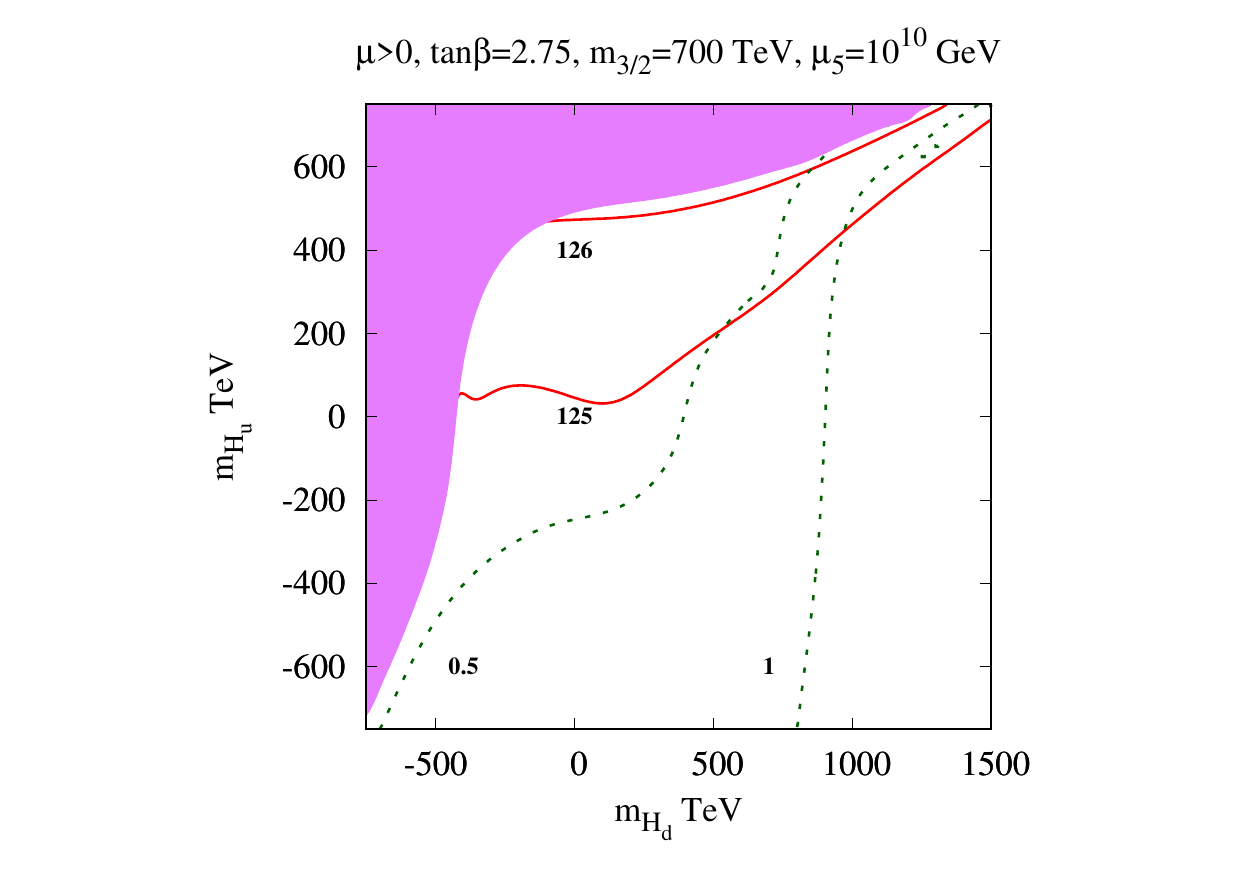}
\end{minipage}

\caption{\label{fig:m1vm2}
The $m_{H_u}^2$ and $m_{H_d}^2$ dependence of the lightest Higgs boson mass in units of GeV (red lines) and the proton lifetime in units of $10^{35}$\,years (green dotted lines).
The blue dashed line is where the Wino masses give the correct thermal relic density for dark matter.
The yellow dashed line is where $c_W = 0$.
The pink shaded region is excluded by the failure of the EWSB.
}

\end{figure}

 Next, we look at the $\tan\beta$ vs $m_{3/2}$ plane for $m_{H_u}^2=0$.  We choose $m_{H_u}^2=0$ for simplicity. However, most other values where $m_{H_u}^2$ is smaller than $m_{3/2}^2$ by a non-trivial amount would work. The advantage of taking $m_{H_u}^2< m_{3/2}^2$ is it restores the freedom in $\tan\beta$. In universal PGM, small values of $\tan\beta$ are needed so that the top Yukawa couplings is large. If the top Yukawa couplings is not large, the radiative corrections to the Higgs soft masses are not large enough to generate radiative EWSB.
 In Fig.\,\ref{fig:tbvsm32mhu0}, we show the plane of $\tan\beta$ versus $m_{3/2}$. The blue dashed line is where the Wino masses give the correct thermal relic density for dark matter.
 Below this line, the dark matter density is less than the experimentally measured value. In the region below this line, the dark matter can be a mixture of the Wino and axion.  As can be seen from the green short dashed line, the proton lifetime, which is labeled in units of $10^{35}$\,years, is quite small in the regions which is preferred by the Higgs mass measurements.  This means future experiments will be able to completely rule out all parameter space shown in this figure.  Furthermore, the constraints on the proton lifetime push us toward smaller $m_{3/2}$ and thus a larger fraction of axion dark matter. Effectively, this model correlates the axion dark matter fraction with proton lifetime.

The last set of figures, Fig.\,\ref{fig:m1vm2}, are for the $m_{H_d}$ versus $m_{H_u}$ plane\footnote{Here, $m_{H_d}$ and $m_{H_u}$ denote $\mathrm{sign}(m_{H_d})|m_{H_d}^2|^{1/2}$ and $\mathrm{sign}(m_{H_u})|m_{H_u}^2|^{1/2}$, respectively. }.
In these figures, we show how the proton lifetime depends on the Higgs soft masses. In the top left figure, we take $m_{3/2}=200$\,TeV.
The red contours are the Higgs mass, the green short dotted lines are the proton lifetime in units of $10^{35}$ years, and the yellow short dashed line is where $c_W=0$. There is no line corresponding to a dark matter density of $0.12$, since the entire plane has a Wino masses which is too small to give a thermal dark matter density of $0.12$. The proton lifetime varies quite slowly across the entire plane. In fact, most of the plane is within reach of upcoming proton decay searches. The pink region along the top and left edge is excluded because the radiative EWSB conditions cannot be met. In the top right figure, we take the same set of parameters except now we take $m_{3/2}=500$\,TeV. the lines are the same as the left figure except now we have a blue long dashed line corresponding to the measure relic density 0.12.  Again, the proton lifetime varies slowly and much of the plane is within reach of upcoming experiments. In the bottom left figure, we take $m_{3/2}=700$ TeV. This figure is similar to the top two figures, except now the dark matter density is too larger over much of the plane. This mean if the universe follows a standard cosmology, we are constrained to live along the edge of the region where electroweak symmetry breaking fails.  The bottom right figure, is the same as the bottom left except it has $\mu<0$.  This drastically affects the dark matter density, since if flips the sign of the threshold correction to the wino coming from integrating out the Higgsino.  This drastically reduced the dark matter density and makes the entire plane have a dark matter density less than 0.12.

\section{Conclusions}
In this paper, we discussed the proton decay for the simplest product group unification based on SU(5)$\times$U(2)$_\mathrm{H}$.
The product group unification is
attractive alternative which
solves the doublet-triplet splitting problem and the dimension-5 proton decay problem by R-symmetry.
By requiring the model be perturbative up to the cutoff scale, we find that the effective GUT scale is considerably smaller than the conventional GUT scale, which roughly corresponds to the scale the MSSM gauge coupling constants unify.
As a result, we find that the minimal setup of the SU(5)$\times$U(2)$_\mathrm{H}$ model has been excluded by the proton decay experiments.

We also showed that a simple extension of the model with SU(5) incomplete multiplets can rectify this problem.
It should be noted that the incomplete multiplets can be achieved in product group unification without fine-tuning.
Although the proton lifetime does not depend on the MSSM spectrum significantly, we demonstrated the parameter dependence by taking the PGM spectrum as an example.
As a result, we found that the proton lifetime in the extended model is in reach of coming experiments like DUNE and Hyper-K, when the mass of the incomplete multiplet is associated with the Peccei-Quinn symmetry.
The dark matter in this model consists of
an admixture of the Wino LSP and the axion.
The axion coupling to QED is enhanced by a factor of three compared with the KSVZ axion model with a GUT complete $\mathbf{5}$, $\mathbf{\bar{5}}$ multiplet.
Therefore,
this scenario can be tested by combining the proton decay searches, the LSP (Wino) searches and the axion searches.

Product group unification models based on SU(5)$\times$U(3)$_\mathrm{H}$ are also possible~\cite{Ibe:2003ys,Izawa:1997he,Yanagida:1994vq,Hisano:1995hc,Hotta:1996pn}. The minimal model is likewise ruled out due to a short dimension-6 proton decay lifetime. Similar to what was done in here, this class of models can be salvaged by the addition of intermediate scale SU(2) doublet fields. However, there is no strong motivation for these intermediate mass doublets. Unlike the SU(5)$\times$U(2)$_\mathrm{H}$ case, where, the intermediate scale mass of the colored triplets is set by PQ breaking scale. Thus, the preferred product group unification models is the one we have considered based on SU(5)$\times$ U(2)$_\mathrm{H}$.

\section*{Acknowledgments}
This work is supported by Grant-in-Aid for Scientific Research from the Ministry of Education, Culture, Sports, Science, and Technology (MEXT), Japan, 17H02878 (M.I. and T.T.Y), 18H05542 (M.I.), 19H05810 (T.T.Y), and by World Premier International Research Center Initiative (WPI), MEXT, Japan. J.E. and T.T.Y. would like to thank IPMU for their hospitality during the completion of this work.

\bibliographystyle{apsrev4-1}
\bibliography{ref}





\end{document}